\documentclass[sigconf, nonacm]{acmart}
\usepackage{amsmath}
\usepackage{graphics}
\usepackage{listings}
\usepackage{appendix}

\AtBeginDocument{%
	\providecommand\BibTeX{{%
			\normalfont B\kern-0.5em{\scshape i\kern-0.25em b}\kern-0.8em\TeX}}}

\begin{document}
	\title{Synchronous Consensus During Incomplete Synchrony}
	
	\author{Ivan Klianev}
	\email{Ivan.Klianev@gmail.com}
	\affiliation{%
	\institution{Transactum Pty Ltd}
	\city{Syndey}
	\country{Australia}
	}
	\begin{abstract}
		
	We present an algorithm for synchronous deterministic Byzantine consensus, tolerant to links failures and links asynchrony. It cares for a class of networks with specific needs, where both safety and liveness are essential, and timely irrevocable consensus has priority over highest throughput. The algorithm operates with redundant delivery of messages via indirect paths of up to 3 hops, aims all correct processes to obtain a coherent view of the system state within a bounded time, and establishes consensus with no need of leader. Consensus involves exchange of $2 * n^{3}$ asymmetrically authenticated messages and tolerates up to < n/2 faulty processes.

	We show that in a consensus system with known members: 1) The existing concepts for delivery over a fraction of links and gossip-based reliable multicast can be extended to also circumvent asynchronous links and thereby convert the reliable delivery into a reliable bounded delivery. 2) A system of synchronous processes with bounded delivery does not need a leader – all correct processes from connected majority derive and propose the same consensus value from atomically consistent individual views on system's state. 3) The required for bounded delivery asymmetric authentication of messages is sufficient for safety of the consensus algorithm.
	
	Key finding: the impossibility of safety and liveness of consensus in partial synchrony \cite{GilbertLynch_2012} is not valid in the entire space between \textit{synchrony} and \textit{asynchrony}.	
	A system of synchronized synchronous processes, which communicate with asymmetrically authenticated messages over a medium susceptible to asynchrony and faults, can operate with: 1) defined tolerance to number of asynchronous and/or faulty links per number of stop-failed and/or Byzantine processes; 2) leaderless algorithm with bounded termination; and 3) conceptually ensured simultaneous safety and bounded liveness. 
		
	\end{abstract}

		

	\maketitle	
	
	\section{Introduction}
	 	
	\subsection{Theoretical Background}	
		
		Deterministic synchronous consensus has conceptual significance for the purposes of atomicity of committed distributed database transactions \cite{MohanStrongFink1985} 
		\cite{HectorGM_EtAl1986}. 
		It is impossible with a large fraction of all communication links being faulty \cite{JimGray78}. 
		With all links being correct and process faults up to a fraction of all processes, it was solved with an algorithm where the decision making takes as input a vector, having an element for each process, loaded with values received from all processes  
		\cite{PeaseShostakLamport80} 
		\cite{LamportShostakPease82}. Under the same limit of tolerated faulty processes, it was also solved with a small fraction of all one-way links being faulty \cite{Dolev1981}. Aspects of synchronous consensus under simultaneous process and link failures were studied and solved in various contexts and scenarios 
		\cite{MeyerPradhan1991} 
		\cite{Hin-SingSiuEtAl1998}
		\cite{Schmid2001}
		\cite{SchmidEtAl2002}
		\cite{SantoroEtAl2007}
		\cite{SchmidEtAl2009}
		\cite{BielyEtAl2011}.
		Yet none of these works considers possible communication asynchrony.
		
		Consensus cannot be solved with a single unannounced failure of a process under \textit{full asynchrony} of processes, communication, and message ordering \cite{FLP_result} and also under full asynchrony of either processes or communication \cite{DolevDworkStockm1987}. Yet the needs of practice compel more attention to communication that is neither fully asynchronous nor fully synchronous. Its name \textit{partial synchrony} \cite{DworkLynchStockm1988} reflects the reality where intervals of full synchrony with unknown duration are followed by intervals of infringed full synchrony. First dealing with it was under the model of \textit{eventual synchrony} \cite{DworkLynchStockm1988}, meaning eventual \textit{in time} synchrony. This model stoically accepts the inevitability of intervals with asynchrony and allows termination only during large enough intervals of full synchrony.
		\cite{DuttaGuerraoui_2002}
		\cite{Alistarh_2008}.
		
		Handling asynchrony in a proactive manner makes more sense. One, and so far the most popular, way of doing that is to give up the reliance on ability of correct processes to obtain coherent view of the system state for computation of consensus value and replace it with reliance on a leader to propose the value \cite{CastroLiskov2002PBFT}. This approach is a tradeoff where vulnerability to stop-fail or Byzantine faults \cite{LamportShostakPease82} of the leader replaces the vulnerability to communication asynchrony. A Byzantine-faulty leader can delay consensus without revealing itself. Hence, with this way of proactive handling of asynchrony, as well as with the passive waiting for its disappearance, consensus is time vulnerable. These outcomes led to conjecture of impossibility to ensure simultaneous safety and liveness of consensus systems under partial synchrony \cite{GilbertLynch_2012}. Leaderless consensus under partial synchrony is considered theoretically possible \cite{BoranSchiper_2010}. Yet at present all deterministic algorithms that solve Byzantine consensus over communication medium susceptible to asynchrony are leader-based 
		\cite{Abraham_EtAl_2020}
		\cite{Buchman_EtAl_2018}
		\cite{Baudet_EtAl_2020}
		\cite{Naor_EtAl_2019}
		\cite{Naor_Keidar_2020}
		\cite{Bravo_EtAl_2020}
		\cite{Stathakopoulou_EtAl_2019}, 
		hence time-vulnerable. 
		
		\textbf{Definition 1} (\textit{Leaderless Consensus Algorithm}): Every process, from a quorum of correct processes with atomically consistent \cite{Herlihy_Wing_1990} views of system's state, proposes the same value for consensus.
	
		Consensus is key for ensured atomic operations across database management systems \cite{Gray_lamport}. These systems implement and use local transactions for consistency of operations and integrity of local data. Distributed transactions across these systems aim the same outcome and also by using local transactions, yet executed under external coordination. It involves use of a protocol from a family of distributed commit protocols 
		\cite{JimGray78} 
		\cite{BruceLindsay79} 
		\cite{LampsonSturgis76}, 
		known for susceptibility to blocking 
		\cite{BernsteinHadzilacosGoodman}
		\cite{WeikumVossen}
		and/or possibility to leave a database system participating in a distributed transaction in inconsistent state. 
		
		\textbf{Definition 2} (\textit{Decentralized Database System}): Data replicated on multiple database management system peers in different trust domains and modified with decisions made by quorum of peers.
		
		Decentralized database systems solve half of the problem 
		solved with distributed transactions – the stretch over distance – with global replication of locally executed transactions. These solutions ensure atomic (linearizable) 
		\cite{Herlihy_Wing_1990} consistency across local systems' data with state machine replication 
		\cite{Lamport_StateMachine} 
		\cite{Schneider_1990} 
		\cite{Lampson_1996}. 
		Users can safely transact across continents, but only within a decentralized database silos. Solving the second half requires non-blocking transactions across the boundaries of decentralized database systems.
		
		For the duration of a distributed transaction, all participants performing a role in it (the transacting parties and the coordinator) operate as a single system. Its partitioning, or spoiled liveness of a participant, normally spoils its liveness and may compromise its safety, regardless of non-compromised safety of all participants, and the transaction cannot fulfill its purpose to ensure consistent outcome. Hence, distributed transaction system's prerequisites for safety are: 1) non-partitioning; and 2) liveness and safety of all participants. Further on in the paper we elaborate on and present a way to tame the root cause for partitioning and spoiled liveness.

	\subsection{Problem Formulation}
		
		Root cause for the present inability of deterministic Byzantine consensus systems to operate with ensured safety and bounded liveness under partial synchrony \textbf{is} the perceived impossibility for bounded delivery during incomplete communication synchrony. 
		
	\subsection{Motivation}
	
		Possibility for deterministic Byzantine consensus systems with ensured safety and bounded liveness in partial synchrony would make feasible distributed transactions between decentralized database systems to operate with consistency resilient to:
		
		- Partitioning of the distributed transaction system; or
		
		- Fraudulently 'lost' liveness by one participant. 
		
		\subsection{Solution Brief}
		A digitally signed \cite{RivestShamirAdleman78} message proves its authenticity to anyone with access to signing party’s public key. So, it might be relayed multiple times before reaching its final destination, in effect as if received directly from the sender. The key idea is to circumvent faulty and/or asynchronous links with epidemic dissemination \cite{BirmanEtAl1999} of messages and attain performance of a system with fault-free and asynchrony-free links. The goal is all correct processes to obtain a coherent view of the system state within a bounded time and establish consensus with no need of leader. From engineering perspective the solution is a trade-off: giving up the max throughput and efficient scalability in exchange for ensured bounded liveness. 
		
	\subsection{Similar Works}
		
		This solution exhibits a major distinction from all previous works. It circumvents asynchronous links with bounded delivery over the links that operate synchronously. No known concept for delivery over a fraction of links considers links asynchrony or tolerates a non-negligible amount of it. Similarly, no known gossip-based concept or algorithm is concerned with bounded delivery.
		
	
		\textbf{Synchronous Consensus Over a Fraction of Links}
	
		Siu \textit{et al} \cite{Hin-SingSiuEtAl1998} presents a concept for fault-tolerant virtual channels in a synchronous deterministic Byzantine consensus system, where processes and links can fail independently and simultaneously. It ensures reliable delivery for the purposes of synchronous consensus over a fraction of links via circumvention of faulty link.
		This concept does not consider asynchrony of links.
	
		Aguilera \textit{et al} \cite{AguileraEtAl2003} \cite{AguileraEtAl2004} demonstrates a case of solved synchronous deterministic Byzantine consensus with a small fraction of all links operating synchronously. It is rather an exception from the normal distribution of asynchrony – one correct process having all inbound and outbound links synchronous. The number of permutations with such outcome shows this case's negligibly small probability.
	
		\textbf{Consensus via Gossip-based Reliable Multicast}	
	
		In a distributed system where every process knows every other, the algorithm of Birman \textit{et al} \cite{BirmanEtAl1999} uses gossips, a.k.a. epidemic dissemination of messages, to deliver to all recipients, even if the sender fails before sending to all. The algorithm tolerates process crash failures and a bounded number of soft failures of links, such as buffer overflown. The algorithm does not consider Byzantine failures, asynchrony of links, or bounded delivery. 
	
		In a distributed system where it is impractical every process to know every other, Eugster \textit{et al} \cite{EugsterEtAl2004} ensures probabilistic message delivery by gossiping with randomly selected partners. This work does not consider Byzantine-faulty processes or bounded delivery during asynchrony of some links. With the same type of distributed system, Guerraoui \textit{et al} \cite{GuerraouiEtAl2019} goes further to ensure the same as \cite{EugsterEtAl2004} under presence of Byzantine-faulty processes. Yet this work also does not consider possible asynchrony of links or bounded delivery under Byzantine omissions / delays.	
		
	\subsection{Contribution}
		
		This paper contributes on the subject matter of consensus in partial synchrony:
		
		1. The algorithm for synchronous deterministic consensus with highest known tolerance to communication asynchrony;
		
		2. The first leaderless consensus algorithm over a communication medium susceptible to asynchrony and faults; 
		
		3. Proof of possibility for deterministic Byzantine consensus systems to operate with ensured safety and bounded liveness.

	\subsection{Content}		
		The rest of the paper is structured as follows: Section 2 contains thorough formal presentation of the concept. Section 3 presents the system model of distributed computing. Section 4 presents the algorithm. Section 5 proves the claims of this paper. Section 6 summarizes the results. Section 7 concludes the paper. Appendix shows assessment of tolerance with simulation and regression analysis.

	
	\section{The Concept}
		
		A digitally signed \cite{RivestShamirAdleman78} message might be relayed multiple times before reaching its final destination, in effect as if received directly from the sender. Thus, in spite of faults and/or asynchrony of some links, a message might reach all recipients via \textit{indirect channels} – virtual aggregations of multi-hop indirect delivery paths.

		\subsection{Direct Channels}
		
		Distributed consensus system $S$ of $N$ processes $(N \geq 3)$ has a full set $C$ of $N*(N-1)$ direct channels $c_{pq}$ $(1 \leq p \leq N, 1 \leq q \leq N, p \neq q)$. Channel $c_{pq}$ represents a one-way link for delivery from process $p$ to process $q$. Channel value 1 denotes \textit{correct and synchronous} link. Value 0 denotes \textit{\textit{faulty or asynchronous}}. Matrix $C$ represents the full set $C$ and also $N$ values representing $c_{pq}$ $(1 \leq p \leq N, 1 \leq q \leq N, p = q)$, i.e. a link of every process to itself, always with value 1:
		
		\begin{equation}
			\label{eq:1}
			C = \begin{Bmatrix}
				c_{11} & c_{12} & \dots & c_{1N} \\
				c_{21} & c_{22} & \dots & c_{2N} \\
				\vdots & \\
				c_{N1} & c_{N2} & \dots & c_{NN}
			\end{Bmatrix}
			\tag{1}
		\end{equation}
		
		
		\subsection{Two-hop Indirect Delivery Paths}
		
		System $S$ has also a full set $C'$ of \textit{two-hop indirect} delivery paths comprising $N*(N-1)*(N-2)$   individual paths, as every direct path has $(N-2)$ two-hop alternatives. A message from process $p$ to process $q$ can be delivered with two hops, the first hop from $p$ to process $r$ and the second hop from process $r$ to $q$. Indirect delivery over two links with bounded delivery is longer but still bounded. A \textit{two-hop indirect} delivery path is denoted in two ways:
		
		- $c'_{pqr} (1 \leq p \leq N, 1 \leq q \leq N, 1 \leq r \leq N, p \neq q,  r \neq p,  r \neq q )$, indicating delivery from process $p$ to process $q$ via process $r$; or
		
		- $c'_{pq}$, indicating delivery over any of the \textit{two-hop indirect} delivery paths from process $p$ to process $q$ represented by a vector of $(N-2$) elements, each representing a vector of 2 elements: $c_{pr}$ and $c_{rq}$. 
		
		Value of $c'_{pqr}$ is 1 if both $c_{pr}$ and $c_{rq}$ have value 1. Value of $c'_{pq}$ is 1 if among the existing $(N-2)$ indirect delivery paths there exists a path where both $c_{pr}$ and $c_{rq}$ have value 1.  
		
		\textbf{Formal}
		
		
		Matrix $C'$ represents a full set of two-hop indirect delivery paths:
		\begin{subequations}
			\begin{equation}
				\label{eq:2}
				C' = \begin{Bmatrix}
					c'_{11} & c'_{12} & \dots & c'_{1N} \\
					c'_{21} & c'_{22} & \dots & c'_{2N} \\
					\vdots & \\
					c'_{N1} & c'_{N2} & \dots & c'_{NN}
				\end{Bmatrix}
				\tag{2}
			\end{equation}
			where value of $c'_{pq} (p = q)$ is always 0 as it has no meaning and where value of $c'_{pq} (p \neq q)$ is defined by vector $C'_{pq}$:
			\begin{equation}
				\label{eq:2.1}
				C'_{pq} =
				\begin{Bmatrix}
					k'_{1} & k'_{2} & \dots & k'_{N-2}
				\end{Bmatrix}
				\tag{2.1}
			\end{equation}
			comprising all two-hop paths from $p$ to $q$, and computed as: 
			\begin{equation}
				\label{eq:2.2}
				c'_{pq}=k'_{1} \lor k'_{2}\lor \dots \lor k'_{N-2}
				\tag{2.2}
			\end{equation}
			where $k'_{i} = c'_{pqr}$, and value of $c'_{pqr}$ is defined by vector $C'_{pqr}$:
			\begin{equation}
				\label{eq:2.3}
				C'_{pqr} =
				\begin{Bmatrix}
					k_{1} & k_{2}
				\end{Bmatrix}
				\tag{2.3}
			\end{equation}
			specifying the links participating in the path, and computed as:
			\begin{equation}
				\label{eq:2.4}
				c'_{pqr} = k_{1} \land k_{2}
				\tag{2.4}
			\end{equation}
		\end{subequations}
		
		
		\subsection{Three-hop Indirect Delivery Paths}
		
		System $S$ has as well a full set $C''$ of \textit{three-hop indirect} delivery paths comprising $N*(N-1)*(N-2)*(N-3)$ paths as every \textit{two-hop indirect} path has $(N-3)$ three-hop alternatives. A message from process $p$ to process $q$ can be delivered with three hops: the first from $p$ to process $r$, the second from $r$ to process $s$, and the third from process $s$ to $q$. A \textit{three-hop indirect} delivery path is denoted also in two different ways:
		
		- $c''_{pqrs} (1 \leq p \leq N, 1 \leq q \leq N, 1 \leq r \leq N, 1 \leq s \leq N, p \neq q,  r \neq p,  r \neq q, s \neq p, s \neq q, s \neq r )$, indicating delivery from $p$ to $q$ via processes $r$ and $s$; or
		
		- $c''_{pq}$, indicating any of the \textit{three-hop indirect} delivery paths from $p$ to $q$, represented by a matrix of $(N-2)*(N-2)$ elements, each representing a vector of 3 elements: $c_{pr}$, $c_{rs}$, and $c_{sq}$.
		
		Value of $c''_{pqrs}$ is 1 if each of its constituent elements $c_{pr}$, $c_{rs}$, and $c_{sq}$ has value 1. Value of $c''_{pq}$ is 1 if among the existing $(N-2)*(N-2)$ indirect delivery paths there exists a path where each of its constituent elements $c_{pr}$, $c_{rs}$, and $c_{sq}$ has value 1. 
		
		
		\textbf{Formal}
		
		
		Matrix $C''$ represents full set of three-hop indirect delivery paths:
		\begin{subequations}
			\begin{equation}
				\label{eq:3}
				C'' = 
				\begin{Bmatrix}
					c''_{11} & c''_{12} & \dots & c''_{1N} \\
					c''_{21} & c''_{22} & \dots & c''_{2N} \\
					\vdots & \\
					c''_{N1} & c''_{N2} & \dots & c''_{NN}
				\end{Bmatrix}
				\tag{3}
			\end{equation}
			where the value of $c''_{pq} (p = q)$ is always 0 as it has no meaning and the value of $c''_{pq} (p \neq q)$ is defined by matrix $C''_{pq}$, where $M=N-2$:
			
			\begin{equation}
				\label{eq:3.1}
				C''_{pq} = 
				\begin{Bmatrix}
					k''_{11} & k''_{22} & \dots & k''_{1M} \\
					k''_{12} & k''_{22} & \dots & k''_{2M} \\
					\vdots & \\
					k''_{M1} & k''_{M2} & \dots & k''_{MM} \\
				\end{Bmatrix}
				\tag{3.1}
			\end{equation}	
			comprising all $M*M$ 3-hop paths from $p$ to $q$, and computed as:
			\begin{align}
				\label{eq:3.2}
				\tag{3.2}
				c''_{pq} =	&k''_{11} \lor k''_{12} \lor \dots \lor k''_{1M} \lor \nonumber \\
				&k''_{21} \lor k''_{22} \lor \dots \lor k''_{2M} \lor \nonumber \\
				&\vdots \nonumber \\
				&k''_{M1} \lor k''_{M2} \lor \dots \lor k''_{MM} \nonumber 
			\end{align}	
			where value of $k''_{ij} = c''_{pqrs}(r=s)$ is 0 as it has no meaning and where value of $k''_{ij} = c''_{pqrs}(r\neq s)$ is defined by vector $C''_{pqrs}$:
			\begin{equation}
				\label{eq:3.3}
				\tag{3.3}
				C''_{pqrs} =
				\begin{Bmatrix}
					k_{1} & k_{2} & k_{3}
				\end{Bmatrix}
			\end{equation}
			specifying the links participating in the path, and computed as:
			\begin{equation}
				\label{eq:3.4}
				\tag{3.4}
				c''_{pqrs} = k_{1} \land k_{2} \land k_{3} 
			\end{equation}
		\end{subequations}
		For instance, value of $c''_{1234} = c_{13} \land c_{34} \land c_{42}$.
		
		
		\subsection{Subsets of Channels}
		
		At any point of time, only a fraction of links in $S$ may be correct and operate synchronously. Yet, $S$ may have $D$ \textit{delivery} subsets $Cd_{i} (1 \leq i \leq D)$, where the channels of every subset $Cd_{i}$ can deliver all messages, directly or indirectly, within the synchrony bound. Given that all $N$ processes in  $S$ are correct, a subset of $C$ \eqref{eq:1} is a \textit{delivery} one if it can deliver a message from a process $p$ to any other process $q$ over $c_{pq}$ $(1 \leq p \leq N, 1 \leq q \leq N, p \neq q)$ \eqref{eq:1} or over $c'_{pq}$ \eqref{eq:2} or over $c''_{pq}$ \eqref{eq:3} of the subset. A \textit{non-delivery} subset is one where no $c_{pq}$ and no $c'_{pq}$ and no $c''_{pq}$ can deliver the message. Existence of at least one \textit{delivery} subset $Cd_{i}$ is the necessary and sufficient condition for delivery of all messages. Hence, the delivery of all messages can be prevented only by a number of faulty direct channels that renders all subsets \textit{non-delivery}.
		
		\subsection{Effect of Faulty Processes}
		
		In a consensus system where any third party can verify authenticity of messages, a Byzantine-faulty process may deceive about its value, but may not tamper any message from other process and send it as authentic \cite{PeaseShostakLamport80}. Its options are limited to failure to send a protocol message to some or all processes, or to deliberately delay its sending \cite{DolevStrong1983}. Hence in a system that relies on indirect delivery, a Byzantine-faulty process can do no more harm than a stop-failed process and consensus safety requires $N=2F+1$ processes, where $F$ is the number of tolerated stop-failed and/or Byzantine-faulty processes.
		The indirect delivery depends on strict execution of an algorithm. A process can spoil an indirect path that relies on its participation. Thus, a faulty process may convert some of the $D$ \textit{delivery} subsets into \textit{non-delivery} subsets. Consequently, system $S$ may have $A$ subsets of channels $Ca_{i} (1 \leq i \leq A)$ called \textit{agreement} subsets, where the channels in every subset $Ca_{i}$ can deliver all messages necessary for reaching an agreement, directly or indirectly. \textit{Agreement} subset of $C$ \eqref{eq:1} is a subset that can synchronously deliver a message from any process $p$ of a group of at least $F+1$ correct processes to any other process $q$ of the same group over $c_{pq}$ $(1 \leq p \leq N, 1 \leq q \leq N, p \neq q)$ \eqref{eq:1} or over $c'_{pq}$ \eqref{eq:2} or over $c''_{pq}$ \eqref{eq:3} of the subset. A \textit{non-agreement} subset is one where neither $c_{pq}$, $c'_{pq}$, or $c''_{pq}$ of the subset can deliver the message synchronously.
		
		\subsection{Virtual Channels for Indirect Delivery}
		
		A set of \textit{Virtual indirect channels} is a construction of links and processes interacting according to a deterministic protocol. Purpose of an indirect channel is to circumvent a faulty or asynchronous link via indirect synchronous delivery. A set of \textit{2-hop indirect channels} operates with one process that helps deliver with two hops. The helper multicasts a message received from sender to $(N-2)$ recipients, thus enacting $(N-2)$ channels; A set of \textit{3-hop indirect channels} operates with $1 +(N-2)$ helper processes. The first helper enacts $(N-2)$ \textit{2-hop indirect channels}. The group of $(N-2)$ second helpers enacts the third hop where each of the second helpers multicasts the received message to $(N-3)$ recipients, on its first receiving.

		
		\section{System Model}
		Critical for the concept is to minimize the chance that all inbound links to a process can together fail or became asynchronous. Failure of a link is typically due to network card's overflown receive buffer. Avoiding it requires hardware with capacity for overlapped receiving of messages with max size from all peers. So, an avalanche of faulty links triggered by a single link fault can be ruled out.
		
		Asynchrony of links is beyond Internet users' ability to tame it. Multiple inbound links to a process inevitably share a router that may become overloaded. Fortunately, TCP congestion avoidance
		\cite{Jacobson_Karels_1988} algorithms use 
		\textit{random early detection}
		\cite{Jacobson1988}
		\cite{FloydJacobson1993}
		\cite{JacobsonEtAl1999} to halve the TCP window of individual flows one-by-one when needed, thereby ruling out simultaneous switch to unbounded delivery.
		
	\subsection{Model Assumptions}
	
		\textbf{Computation}:
		
		- \textit{Deterministic processes}. Processes are implemented as computer code with state-defined deterministic execution on pre-selected CPU cores to run process threads.
		
		- \textit{Deterministic protocol}. Processes communicate with exchange of messages, which follows a deterministic protocol.
	
		\textbf{Communication} and \textbf{messages}:
	
		- \textit{Synchronous with limited exceptions}. Messages are normally delivered within a known bounded amount of time.
	 
		- \textit{Signed}. Messages are signed with third party verifiable digital signatures, i.e. able to prove its authenticity to every process.
	
		\textbf{Processors} and \textbf{processes}:
	
		- \textit{Synchronous}. Processes execute on CPUs with a fixed upper bound on how faster one complete a processing cycle than another and on the difference in cache size and cache latencies.
	
		- \textit{Synchronized}. Processes start a consensus round with a max lag bound to less than $RTTB/2$ (round trip transmission bound), which is $MDTB$ (message delivery time bound).
	
		- \textit{No clocks drift}. Synchronized processes stay synchronized via periodic re-synchronization of clocks.
		
		- \textit{Known public keys}. Every process knows the public key of every other process' private key for signature. 
	
		\textbf{Communication failures and asynchrony}:
	
		- \textit{Uncorrelated failures}. A fault in one link is independent from a fault in another link, i.e. a failure in an inbound link cannot trigger an avalanche of inbound link failures.
		
		- \textit{Uncorrelated asynchrony}. Asynchronous operation of one link is independent from asynchronous operation of another link, i.e. an asynchronous inbound link cannot trigger an avalanche of asynchronous inbound links.
	
		\textbf{ Process failures}:
	
		-\textit{ Stop-Fail}. Processes execute correctly, but can stop at any time and once stopped, cannot restart.
	
		- \textit{Bizantine}. Arbitrary behavior expressed as:
	
		- (1) \textit{Omission} to send to all recipients.
	
		- (2) \textit{Delay} before sending to one or more recipients.
	
		- (2) \textit{Deceit} in regard to message content.
	
	\subsection{The System}
		The model considers a consensus system $S$ comprising $N$ deterministic \textit{processes} $(N \geq 3, N \geq 2F+1)$, where $F$ is the tolerated number of faulty processes. Each process $P$ is uniquely identified within $S$ with a system index $I (1 \leq I \leq N)$. The processes are connected by point-to-point one-way \textit{links}. Every pair of processes is connected with a pair of links $c_{IsIr}$, where $Is$ identifies the sender process and $Ir$ identifies the receiver process. Existence of links where $Is=Ir$, i.e. from every process to itself, is assumed for simpler presentation.
		
		Every process $P$ has special memory locations: \textit{input register} $R^{In}$ and \textit{output register} $R^{Out}$. Register $R^{In}$ can contain up to $x$ bytes in size. Register $R^{Out}$ is a vector of $N$ elements $R^{In}$:
		\begin{subequations}
		\begin{equation}
		\label{eq:4}
		\tag{4}
		R^{Out} =
		\begin{Bmatrix}
			R^{In}_{1} & R^{In}_{2} & \dots & R^{In}_{N}
		\end{Bmatrix}
		\end{equation}		
		
		Every process $P$ also has an unbound amount of internal storage and an allocated in it vector $E$ of $N$ \textit{extended} digital signatures: 
		\begin{equation}
		\label{eq:4.1}
		\tag{4.1}
		E =
		\begin{Bmatrix}
			e_{1} & e_{2} & \dots & e_{N}
		\end{Bmatrix}
		\end{equation}	
		\end{subequations} 
		where $e_{i}$ is extended digital signature over $ R^{In}_{i} $ in form of a tuple of 2 elements: 1) $ h_{i} $ is output of execution of the $ Hash $ function used in $S$ over $ R^{In}_{i} $; and 2) $ s_{i} $ is the signature of process $ P_{i} $ over $ h_{i} $. 
	
		The values in $R^{In}$, $R^{Out}$, and the internal storage of the processes of $S$ define the internal state of $S$. Initial states prescribe fixed starting value for all; in particular $R^{In}$ and $R^{Out}$ start with value 0 of every byte. States in which $R^{Out}$ contains $\geq ( F + 1 )$ elements with values $\neq 0$ are distinguished as decision states. 
		
		Each process $P$ acts deterministically according to a transition function $\Phi$, which cannot change the value of $R^{Out}$ once the process has reached a decision state; i.e., $R^{Out}$ is 'write-once'. System $S$ is specified by the function $\Phi$ and the values in $R^{In}$ of every process.
		
		Processes communicate with signed messages. A \textit{message} is tuple $(T, Is, Id, Io, m, Do)$, where $ T $ is a constant specifying message type, $ Is $ is index of the source process, $ Id $ is index of the destination process, $ Io $ is index of the originator process, $ m $ is 'message value' from a fixed universe $ M $, and $ Do $ is digital signature of the originator process over $ m $. Where $Io \neq Is$, this indicates that the message $ m $ sent by process $ P_{Is} $ contains the message and the digital signature that $ P_{Is} $ has received from $ P_{Io} $.

		Communication within a consensus round happens in two phases. During Phase One each process $ P_{Is} $ transmits to every process $ P_{Id} (1 \leq Id \leq N, Id \neq Is)$, the value in its input register $ R^{In}_{Is}$ and its signature over the value in $ R^{In}_{Is} $. During Phase Two each process $ P_{Is} $ transmits the content of its $E_{Is}$ \eqref{eq:4.1} and its signature over $E_{Is}$ to every process $ P_{Id} (1 \leq Id \leq N, Id \neq Ips)$. 		
		
		At the Phase One beginning, state of $S$ is defined by vector V1:
		\begin{subequations}
		\begin{equation}
		\label{eq:5}
		\tag{5}
		V1 =
		\begin{Bmatrix}
			v1_{1} & v1_{2} & \dots & v1_{N}
		\end{Bmatrix}
		\end{equation}

		An element $v1_{i} (1 \leq i \leq N)$ contains the value $R^{In}_{i}$ of process $P_{i}$ and its signature $s_{i}$ over $R^{In}_{i}$. During Phase One each process $P_{i}$ sends $v1_{i}$ to every process $P_{j}(1 \leq j \leq N, j \neq i)$ over $c_{ij}$ \eqref{eq:1}. 
		
		Matrix $D1$ presents what was delivered by $S$ during Phase One. An element $d1_{ij}$ is the value of $v1_{i}$ as delivered over $c_{ij}$, i.e. where value of  $c_{ij}$ is $0$, indicating a faulty direct channel, the delivered value $d1_{ij}$ is $0$; otherwise $d1_{ij} = v1_{i}$:
		\begin{equation}
			\label{eq:5.1}
			\tag{5.1}
		D1 = \begin{Bmatrix}
			d1_{11} & d1_{12} & \dots & d1_{1N} \\
			d1_{21} & d1_{22} & \dots & d1_{2N} \\
			\vdots & \\
			d1_{N1} & d1_{N2} & \dots & d1_{NN} \\
		\end{Bmatrix}
		\end{equation}
	
 		On receiving $d1_{ij} \neq 0$ from process $P_{i}$, process $P_{j}$ executes the system's \textit{ Hash} function over $R^{In}_{i}$ component of $d1_{ij}$, writes the outcome as $h_{i}$ component of element $e_{i}$ of its vector $E_{j}$, uses the signature $s_{i}$ component of $d1_{ij}$ to verify the signature, writes $s_{i}$ as the $s_{i}$ component of element $e_{i}$ of its vector $E_{j}$, and in case of correctly verified signature writes $R^{In}_{i}$ in its vector $R^{Out}{j}$ \eqref{eq:4}.

		On completion of Phase One, every process $P_{j}$ has formed a vector of extended digital signatures $E_{j}$:
		\begin{equation}
			\label{eq:5.2}
			\tag{5.2}
		E_{j} =
		\begin{Bmatrix}
			e_{1j} & e_{2j} & \dots & e_{Nj}
		\end{Bmatrix}
		\end{equation}	
		from  vector $D1_{j}$:
		\begin{equation}
			\label{eq:5.3}
			\tag{5.3}
		D1_{j} = 
		\begin{Bmatrix}
			d1_{1j} \\
			d1_{2j} \\
			\vdots  \\
			d1_{Nj} \\
		\end{Bmatrix}
		\end{equation}
		\end{subequations}
		Next, $P_{j}$ forms $v2_{j}$ containing the vector of extended digital signatures $E_{j}$ and a digital signature of $P_{j}$ over the content of vector $E_{j}$.
		
		At the Phase Two beginning, state of $S$ is defined by vector V2:
		\begin{subequations}
		\begin{equation}
			\label{eq:6}
			\tag{6}
		V2 =
		\begin{Bmatrix}
			v2_{1} & v2_{2} & \dots & v2_{N}
		\end{Bmatrix}
		\end{equation}	
		
		During Phase Two each process $P_{i}$ sends $v2_{i}$ to every other process $P_{j}(1 \leq j \leq N, j \neq i)$ over $c_{ij}$. Matrix $D2$ presents the outcome of Phase Two. An element $d2_{ij}$ is $v2_{i}$ as delivered over $c_{ij}$:
		\begin{equation}
		\label{eq:6.1}
		\tag{6.1}
		D2 = \begin{Bmatrix}
			d2_{11} & d2_{12} & \dots & d2_{1N} \\
			d2_{21} & d2_{22} & \dots & d2_{2N} \\
			\vdots & \\
			d2_{N1} & d2_{N2} & \dots & d2_{NN} \\
		\end{Bmatrix}
		\end{equation}
		\end{subequations}
		
		In case of no process or link failures, all $d2_{ij}$ contain the same value. Process $P_{j}$
		having received $(N-1)$ vectors $d2_{ij}$ that have the same value as the value $d2_{jj}$ that $P_{j}$ received from itself, knows that every other process $P_{i}$ has received exactly the same values. So $P_{j}$ obtains the value upon which an agreement has been reached with execution of a pre-defined deterministic function that takes as input the content of vector $R^{Out}_{j}$.

	\subsection{Handling Faults and Asynchrony}

		The model assumes signed messages able to prove its authenticity to any process of the system. So, a Byzantine process may deceive about its value, but may not alter values received from other processes and send the altered values without revealing itself as malicious \cite{PeaseShostakLamport80}. Thus the model limits the undetected behavior of Byzantine processes to just failure to send a protocol message to some or all processes or deliberately delay the sending \cite{DolevStrong1983}.	

		\textbf{Handling Byzantine Deceit}

		During Phase Two the consensus algorithm exchanges extended signatures only. The model has to ensure that Byzantine consensus can still be solved with $N=2F+1$ processes. A Byzantine-faulty process $P_{i}$  may send different versions of $v1_{i}$ to $P_{j}$ and  $P_{k}$ $(1 \leq j \leq N, 1 \leq k \leq N, j \neq i, k \neq i, j \neq k)$, thus deceiving about own value. If $R^{In}_{i}$ and $s_{i}$ do not match, $v1_{i}$ is ignored; otherwise $P_{j}$ includes $e_{i}$ in vector $E_{j}$ and $P_{k}$ includes $e_{i}$ in vector $E_{k}$. 
		
		On completion of Phase Two, if vectors $E_{j}$ and $E_{k}$ distributed to all processes contain different versions of $e_{i}$, then $P_{i}$ reveals itself as Byzantine-faulty and all correct processes drop $R^{In}_{i}$ from the consensus vector $R^{Out}$. Sufficient is to verify that the different $h_{i}$ of each version is correctly signed with the different $s_{i}$ of its version. Hence system $S$ reveals the deceit and terminates consensus rounds with no need of more than $F+1$ correct processes.
		
		\textit{During Phase One}: A Byzantine-faulty process $P_{i}$ may send more than one value as its $R^{In}_{i}$ value during Phase One. These values may be correctly or incorrectly signed. Process $P_{i}$ is supposed to send value $v1_{i}$ to every process $P_{j} (1 \leq j \leq N, j \neq i)$ over $c_{ij}$. Yet the value $v1_{i}$ that process  $P_{i}$ sends is not the same to every $P_{j}$:
		\begin{subequations}
		\begin{equation}
		\label{eq:7.1}
		\tag{7.1}
		v1_{i} =
		\begin{Bmatrix}
			R^{In}_{i} & s_{i}
		\end{Bmatrix}
		\end{equation}
	
		Process $P_{i}$  may send different versions of $v1_{i}$ to $P_{j} (1 \leq j \leq N, j \neq i)$ and $P_{k} (1 \leq k \leq N, k \neq i, k \neq j)$. Processes $P_{j}$ and $P_{k}$ execute $ Hash $ function over $R^{In}_{i}$ to produce $h_{i}$ and verify correctness of the received signature $s_{i}$. If $R^{In}_{i}$ and $s_{i}$ do not match, $v1_{i}$ is ignored; otherwise $P_{j}$ includes $h_{i}$ and $s_{i}$ as element $e_{i}$ in vector $E_{j}$ and  $P_{k}$ includes $h_{i}$ and $s_{i}$ as element $e_{i}$ in vector $E_{k}$. During Phase Two, vectors $E_{j}$ and $E_{k}$ containing different versions of $h_{i}$  will be distributed to all processes, thereby revealing $P_{i}$ as Byzantine-faulty. So, all correct processes $P_{j} (1 \leq j \leq N, j \neq i)$ will drop $R^{In}_{i}$ from their vector $R^{Out}_{j}$.

		\textit{During Phase Two}: Process $P_{i}$ may deceive about value $v1_{j}$ it has received from process $P_{j} (1 \leq j \leq N, j \neq i)$ during Phase One. Value $v2_{i}$ process $P_{i}$ sends during Phase One contains $E_{i}$ and signature of $P_{i}$ over $E_{i}$:  
		\begin{equation}
		\label{eq:7.2}
		\tag{7.2}
		E_{i} =
		\begin{Bmatrix}
			e_{1i} & e_{2i} & \dots & e_{Ni}
		\end{Bmatrix}
		\end{equation}
		\end{subequations}
		Yet the value $v2_{i}$ sent to at least one process $P_{k} (1 \leq k \leq N, k \neq i)$ contains at least one value $e_{ji}$ that does not reflect correctly the value $v1_{j}$ that $P_{i}$ has received from process $P_{j} (1 \leq j \leq N, j \neq i, j \neq k)$. This deceit cannot be successful: a $h_{j}$ incorrectly reflecting $R^{In}_{j}$ has no chance to match the signature $s_{j}$ of process $P_{j}$.
	
		Process $P_{i}$ may try to deceive about the value received from process $P_{k} (1 \leq k \leq N, k \neq i)$ by including a false $h_{k}$ in its vector $E_{i}$. However, $P_{i}$ cannot produce a false signature $s_{k}$ that matches with the false $h_{k}$. Matching signature can be produced only by process $P_{k}$. Whatever values $h_{k}$ and $s_{k}$ process $P_{i}$ sends, the non-matching between these values will be easily detected by the recipient processes. So, process $P_{i}$ cannot prevent consensus, but reveals itself as Byzantine-faulty. 

		\textbf{Handling Link Failure or Asynchrony}

		Link failure is an arbitrary behavior of a one-way link $c_{ij}$ between two processes $P_{i} (1 \leq i \leq N)$ and $P_{j} (1 \leq j \leq N, j \neq i)$ that prevents correct transmission of messages from the sender process $P_{i}$ to the receiver process $P_{j}$. Asynchrony of the one-way link $c_{ij}$ is arbitrary behavior in regard to delivery time of the communication medium between processes $P_{i}$ and $P_{j}$. Link $c_{ij}$ is represented in matrix $C$ \eqref{eq:1} with value 1 if it operates correctly and synchronously. It has value 0 if it does not deliver within the synchrony bounds.
		
		The model does not distinguish faulty from asynchronous link and does not need to. It handles both with indirect delivery. It does not need the receiving end of a link to distinguish a faulty/asynchronous link from possible process failure on the sending end. 
		
		The model prescribes that in addition to the protocol rules, a correct process $P_{i}(1 \leq i \leq N)$ must follow one more rule. On first receiving of a signed message $mP_{j}$, directly from the signatory  $P_{j} (1 \leq j \leq N, j \neq i)$ or indirectly, $P_{i}$ multicasts $mP_{j}$ to every $P_{k} (1 \leq k \leq N, k \neq i, k \neq j)$ with the signature of $P_{j}$ over $mP_{j}$. 
		
		This rule ensures (where possible) that faulty or asynchronous links will be circumvented with 2-hop or 3-hop indirect delivery. It enacts attempts to deliver message  $mP_{j}$ via different 2-hop indirect channels \eqref{eq:2} and possible attempts to deliver  $mP_{j}$ via different 3-hop indirect channels \eqref{eq:3}.
		
 		\vspace{\baselineskip}
		\textbf{Stop-Failed Process}

		A stop-failed process is one that executes correctly but stops at any time and once stopped, cannot restart. A stop-failed process $P_{i} (1 \leq i \leq N)$ cannot send any $v1_{i}$ or $v2_{i}$ messages to any other process $P_{j} (1 \leq j \leq N, j \neq i)$ over $c_{ij}$ and cannot receive any $v1_{j}$ or $v2_{j}$ messages from any other process $P_{j}$ over $c_{ji}$. The effect of a stop-failed process $P_{i}$ on the functioning of system S is the same as:
		 
		- Simultaneous failure/asynchrony of all links where $P_{i}$ is sender, represented by the entire row with index $i$ of matrix $C$ \eqref{eq:1}, i.e. every element $c_{ij} (1 \leq j \leq N)$ in matrix $C$ having value 0; and
		
		- Simultaneous failure/asynchrony of all links  where $P_{i}$ is receiver, represented by the entire column with index $i$ of matrix $C$ \eqref{eq:1}, i.e. every element  $c_{ji} (1 \leq j \leq N)$ in matrix $C$ having value 0.
		
		\textit{Note}: Process failure presented as link failures provides a common denominator for assessment of the combined effect of simultaneous process failures and link failures/asynchrony and the effect of processes' possible Byzantine behavior in its variety and scale.		
				
		The model does not distinguish whether a missing message is caused by a process stop-failure or anything else. Yet, it takes into consideration that a stop-failed process spoils all 2-hop and 3-hop indirect channels that depend on its active participation. Stop-failed sender during multicast has the effect of Byzantine omission.		
				
		System $S$ needs just $F+$1 correct processes to be able terminate consensus in spite of $F$ stop-failed processes, as a consequence of the eliminated chance for successful Byzantine deceit \cite{PeaseShostakLamport80} 
		\cite{DolevStrong1983}. 
		With this model, correctness of the above statement is subject to additional condition: As long as the $F$ stop-failed processes do not invalidate all indirect channels allowing a correct process to have two-way communication with the rest of the $F$ correct processes.
		
		\textbf{Handling Byzantine Omission}

		Every process $P_{i} (1 \leq i \leq N)$ is required by consensus protocol to send each protocol message to every other process $P_{j} (1 \leq j \leq N, j \neq i)$ over $c_{ij}$. A Byzantine-faulty $P_{i}$ may omit sending to $P_{j}$. Omission by $P_{i}$ to send to  $P_{j}$ over $c_{ij}$ has exactly the same effect on functioning of system $S$ as the event of failure of link $c_{ij}$ and is handled in the same way.
				
		Process $P_{i}$ can omit sending to any process $P_{j}$ without being detected. Sending to at least one process of a group of $F+1$ correct processes with two-way channels between themselves allows the model to tolerate its omissions. Omitted sending to all of the group of $F+1$ correct processes has the effect of a stop-failed process.
		
		\textbf{Handling Byzantine Delay}

		Every process $P_{i}(1 \leq i \leq N)$ is required by consensus protocol to send each protocol message to every process $P_{j} (1 \leq j \leq N, j \neq i)$ over $c_{ij}$ with no delay. A delayed sending by $P_{i}$ to a process $P_{j}$ has the same effect on functioning of consensus system $S$ as sending over asynchronous link, handled exactly as a Byzantine omission.
		
	\subsection{Correctness of Consensus}	
		Consensus requires existence of a group $G$, which comprises  $\geq F+1$ correct processes communicating via direct or indirect channels with bounded delivery.

		\textbf{Correctness Conditions}
		
		\textit{Agreement}: Every two processes $P_{i}$ and $P_{j}$ from $G$ compute the same value for their output registers $R^{Out}_{i}$ and $R^{Out}_{j}$.

		\textit{Termination}: All processes in $G$ decide within 4 RTTB intervals.
		

	\section{The Algorithm}
		
	\subsection{Synchronization}
		The system model assumes synchronized processes
		\cite{HalpernEtAl1984}
		\cite{Lamport_EtAl_1984}
		\cite{Dolev_EtAl_1984}
		\cite{Lundelius_Lynch_1984}
		\cite{Schneider_1986}, 
		with max lag bounded to less than half RTTB (round trip transmission bound). Sufficient condition for ensuring the assumed max lag is every process to have access to a Time server with embedded GPS receiver utilizing atomic-clock synchronization systems 
		\cite{DanaPenrod1990}
		\cite{AllanEtAl1997}
		\cite{MichaelLombardi2010}.
		The processes of a system with synchronized clocks need only to fix the start time of the first consensus round and then periodically re-synchronize the clocks. The last process joining the system broadcasts a signed message setting the start time. Every recipient of that message, on first direct or indirect receiving also broadcasts it together with its original signature. 
		
	\subsection{Flexible Time-bound}
		Public networks like the Internet operate with fluctuating hourly workload over some backbone elements, causing large fluctuations of the normal delivery time. Hence, flexible time-bound for delivery is a precondition for efficiency in ensuring synchrony. The issue is not essential for this paper, so we only acknowledge it briefly.
		
		Backbone routers' congestion control algorithms use \textit{random early detection}
		\cite{Jacobson1988}
		\cite{FloydJacobson1993}
		\cite{JacobsonEtAl1999}
		to prevent buffers overflow by reducing transmission speed of individual flows, possibly multiple times below the maximum. This way, Transmission Control Protocol (TCP) allows routers to handle surges of traffic and avoid channels partitioning. Under partially synchronous communication and the consensus model of \textit{eventual synchrony}, execution might halt as a result of a randomly delayed synchronous flow, seen as asynchrony.
		
		One way of dealing with such random asynchrony is to switch to a larger time-bound for message delivery. Its outcome would be similar to that of routers operating with a fraction of their maximum throughput, yet being unable to eliminate the possibility for random asynchrony. More efficient is a feature similar to TCP congestion window. The consensus algorithm detects early signs of overloaded channels, capable to halt the consensus after further deterioration, and modifies the time-bound for delivery only when needed.

	\subsection{Handling Faulty/Asynchronous Links} 
		
		The algorithm operates in a deterministic data-driven manner by responding to events from a pair of event categories in the sequence of appearing: expiry of time intervals and receiving peer messages.

		\lstset{language=C++,
				basicstyle=\footnotesize}
		\begin{lstlisting}[caption=Processing Timeout Events in C++]
void OnTimeout( BYTE Event )
{
switch(Event)
	{
case START1 :	// Start Phase One
	BroadcastMssgPhaseOne() ;
	break;
case START2 :	// Start Phase Two
	BroadcastMssgPhaseTwo() ;
	break;		
	}
}	
		\end{lstlisting} 
	
		\vspace{\baselineskip}
		\vspace{\baselineskip}
		
		Listing 1 presents the algorithm's part that maintains synchronization between consensus rounds of process $P_{i}(1 \leq i \leq N)$ in system $S (N \geq 3, N \geq 2F+1)$ with the rest of processes $P_{j} (1 \leq j \leq N, j \neq i)$. Function OnTimeout() responds to 2 timeout events: 
	
		- START1 prompts process $P_{i}$ to broadcast its own MSGPHASE1 message to all processes $P_{j} (1 \leq j \leq N, j \neq i)$.
		
		- START2 prompts process $P_{i}$ to broadcast its own MSGPHASE2 message to all processes $P_{j} (1 \leq j \leq N, j \neq i)$.

		\begin{lstlisting}[caption=Processing Received Messages in C++]
void OnRcvdMSSG( MSSG* pMSSG )
{
switch( pMSSG->m_Mssg_Type )
	{
case MSGPHASE1 : // Rcvd a Direct  Mssg Phase One
case RTRNSMTD1 : // Rcvd a Relayed Mssg Phase One
	if( HasMssgBeenRcvd( pMSSG ) break;
	ProcessMssgPhaseOne( pMSSG ) ;
	RetrnsmMssgPhaseOne( pMSSG ) ;
	break;
case MSGPHASE2 : // Rcvd a Direct  Mssg Phase Two
case RTRNSMTD2 : // Rcvd a Relayed Mssg Phase Two
	if( HasMssgBeenRcvd( pMSSG ) break;
	ProcessMssgPhaseTwo( pMSSG ) ;
	RetrnsmMssgPhaseTwo( pMSSG ) ;
	break;			
	}
}			
		\end{lstlisting}
		
		Listing 2 presents the algorithm's part that handles and responds the messages that process $P_{i}(1 \leq i \leq N)$ receives from the rest of processes $P_{j} (1 \leq j \leq N, j \neq i)$. Function OnRcvdMSSG() responds to 4 types of messages:
		
		- MSGPHASE1 is a message that process $P_{i}(1 \leq i \leq N)$ receives directly from every process $P_{j} (1 \leq j \leq N, j \neq i)$. On first receiving, directly or via RTRNSMTD1, $P_{i}$ verifies the signature, records the message in its internal storage, and multicasts it as RTRNSMTD1 message to every process $P_{k} (1 \leq k \leq N, k \neq i, k \neq j)$.
		
		- RTRNSMTD1 is a message signed by $P_{j} (1 \leq j \leq N, j \neq i)$ and received by $P_{i}$ sent by $P_{k} (1 \leq k \leq N, k \neq i, k \neq j)$ in response to MSGPHASE1 from  $P_{j}$.
		
		- MSGPHASE2 is a message that process $P_{i}(1 \leq i \leq N)$ receives directly from every process $P_{j} (1 \leq j \leq N, j \neq i)$. On first receiving, directly or via RTRNSMTD2, $P_{i}$ verifies the signature, records the message in its internal storage, and multicasts it as RTRNSMTD2 message to every process $P_{k} (1 \leq k \leq N, k \neq i, k \neq j)$.
		
		- RTRNSMTD2 is a message signed by $P_{j} (1 \leq j \leq N, j \neq i)$ and received by $P_{i}$ sent by $P_{k} (1 \leq k \leq N, k \neq i, k \neq j)$ in response to MSGPHASE2 from  $P_{j}$.
		
	\begin{figure}[h]
		\scalebox{0.22}
		{\includegraphics{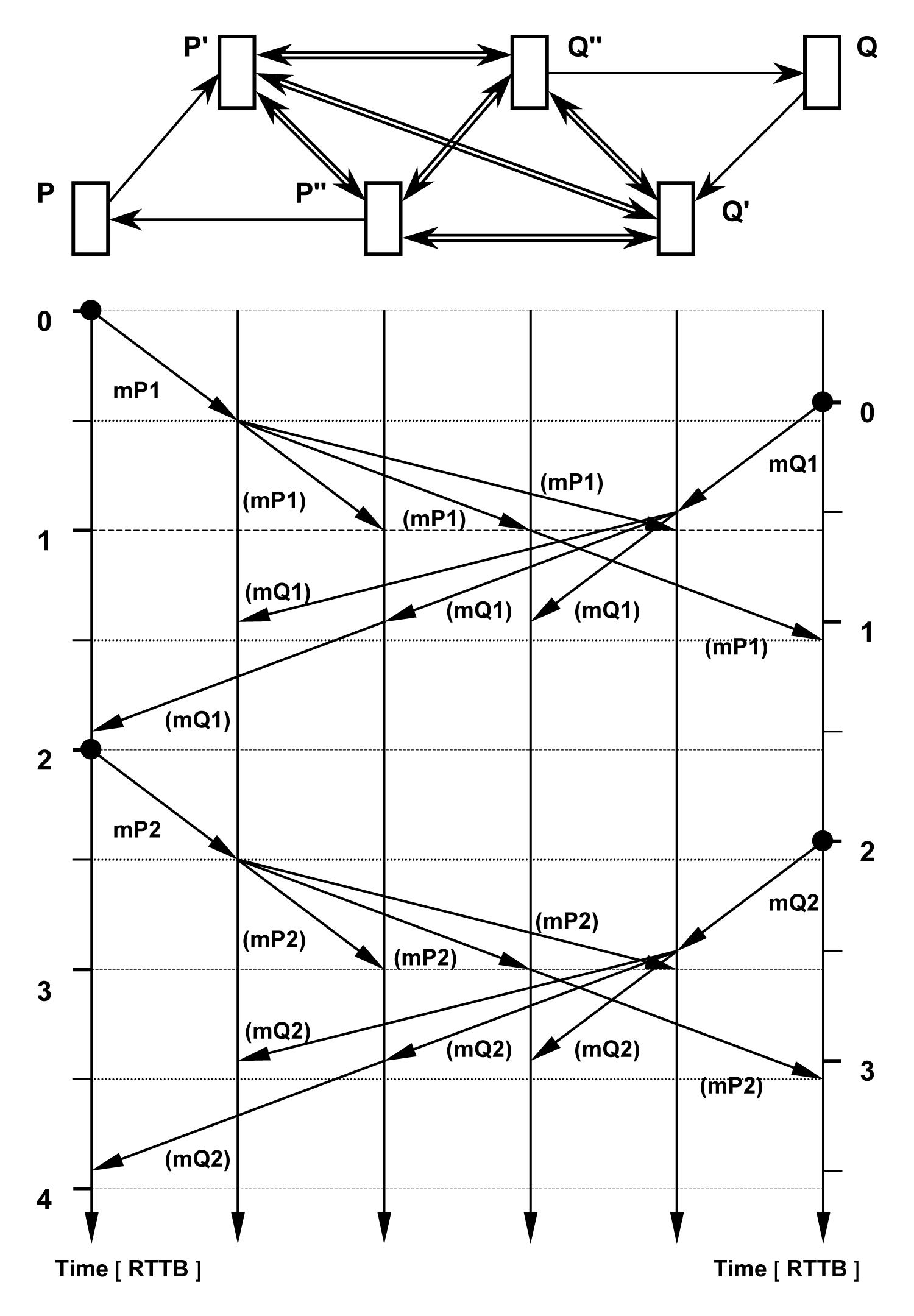}}
		\caption{Communication via 3-hop Indirect Channels}
	\end{figure} 		
	\subsection{Algorithm Illustration}	
		
		Figure 1 is a diagram that illustrates the algorithm operations in a worst-case scenario. On the diagram, all processes are correct  and all presented links are synchronous. The model assumes that the processes start a round with max time lag bound to less than 1/2 RTTB interval. In the presented worst-case scenario:
		
		- delivery of messages over the presented links takes the max allowed time of 1/2 RTTB;
		
		- process $Q$ starts a consensus round after process $P$, lagging with the highest tolerated lag of nearly 1/2 RTTB interval;
		
		- $P$ and $Q$ deliver messages between themselves only via 3-hop indirect channels with the help of processes $P'$, $P''$, $Q'$, and $Q''$.
		           
		On the diagram:
		           
		- $mP1$ denotes MSGPHASE1 message issued by process $P$;
		
		- $(mP1)$ denotes relayed delivery of $mP1$ as RETRANSM1;
		
		- $mQ1$ denotes MSGPHASE1 message issued by process $Q$;
		
		- $(mQ1)$ denotes relayed delivery of $mQ1$ as RETRANSM1.
   
   		The execution sequence for delivery of MSGPHASE1 messages:
 		
 		1. At its point of time 0, $P$ broadcasts $mP1$, which reaches $P'$;
 		
 		2. At its point of time 0, $Q$ broadcasts $mQ1$, which reaches $Q'$;
 		
 		3. Receiving $mP1$, $P'$ multicasts $(mP1)$, which reaches $Q''$;
 		
 		4. Receiving $mQ1$, $Q'$ multicasts $(mQ1)$, which reaches $P''$;
 		
 		5. Receiving $(mP1)$, $Q''$ multicasts $(mP1)$, which reaches $Q$;
 		
 		6. Receiving $(mQ1)$, $P''$ multicasts $(mQ1)$, which reaches $P$;
 		
 		7. $Q$ receives $(mP1)$ before its point of time 2 RTTB;
 		
 		8. $P$ receives $(mQ1)$ before its point of time 2 RTTB.
  		
  		The delivery of MSGPHASE2 messages follows exactly the same patterns.
    
   	\vspace{\baselineskip}
    
    \section{Proofs}
    
 		\subsection{Synchronous Consensus}
 		We will show that under the model assumptions the algorithm operates with synchrony, as illustrated on Figure 1 where processes $P$, $P'$, $P''$, $Q$, $Q'$, and $Q''$ complete their individual consensus rounds before expiry of 4 RTTB on their individual clocks.
 		
 	 	\textbf{Theorem 1}: \textit{Synchronous consensus during incomplete synchrony.}
 	 
 	 	A system of synchronized (less than 1/2 RTTB lag) synchronous processes operating with a 3-hop delivery algorithm, where in a group $G$ of $F+1$ correct processes every process has a synchronous link (1/2 RTTB delivery bound) or an indirect channel over synchronous links to and from every other process in $G$, terminates within 4 RTTB intervals regardless of possibility all other links to be asynchronous.

 		\textbf{Proof} 
 		
 		The trivial case is when $F+1$ correct processes are directly connected to each other via $(F+1)*F$ synchronous links. The proof must show that with a fraction of these links, allowing every pair of processes a direct or indirect bidirectional communication, in any combination of direct and 2-hop and 3-hop indirect channels, each of the $F+1$ correct process terminates within 4 RTTB intervals. 
	
		2-hop indirect channels take up to 1 RTTB interval to deliver a message. 3-hop indirect channels take up to 1.5 RTTB interval to deliver a message. In the worst-case scenario, the first process that starts a round receives a message from every other process before expiry of 2 RTTM on its clock, as 1/2 RTTB lag + 1.5 RTTB delivery = 2 RTTM. All other processes receive a message from every other process within a shorter time on their individual clocks. Hence, everyone of the $F+1$ correct processes in $G$  receives the messages necessary and sufficient for termination within 4 RTTB intervals.
		
	\subsection{Highest Tolerance to Asynchronous Links}	
		Gossiping (epidemic dissemination) is typically used to ensure message delivery under the weakest needs of correct links
		\cite{BirmanEtAl1999}
		\cite{EugsterEtAl2004}
		\cite{GuerraouiEtAl2019}.
		Our algorithm implements gossiping for tolerance to faulty and/or asynchronous links. It does so with up to 3-hop indirect delivery. Any additional hop increases the tolerance but delays termination. 
		
		\textbf{Lemma 1}: \textit{Epidemic dissemination of messages}.
		
		On existence of a path between sender and recipient of a message, an algorithm for epidemic dissemination will deliver the message.
		
		\textbf{Proof}: In an epidemic algorithm every process of a system forwards the received messages over its output links. Thus, the algorithm utilizes all existing in the system multi-hop delivery routes. 
		
		\textbf{Lemma 2}: \textit{Consensus with epidemic dissemination of messages}.
		
		A synchronous  consensus algorithm implementing epidemic dissemination of messages ensures highest tolerance to faulty links.
		
		\textbf{Proof}: It follows from Lemma 1.
		
		\textbf{Theorem 2}:	\textit{Synchronous consensus with tolerance to asynchrony.}
		
		Epidemic dissemination of messages lets synchronous consensus algorithm terminate under highest number of asynchronous links.
		
		\textbf{Proof}
		
		Consider a synchronous consensus algorithm with epidemic dissemination and a system with $L$ one-way links, where either only faulty or only asynchronous links prevent delivery. Both cases are handled in a way taking longer yet bounded time. In the case with only faulty links, let the smallest sufficient for consensus set of messages $M$ is delivered over one of the smallest sets of correct links $S$ allowing delivery [Lemma 1], i.e. with highest tolerated faulty links $F$ [Lemma 2], $F=L-S$. In the case with only asynchronous links, set $S$ delivers the same set $M$, thus tolerating $A$ asynchronous links, where $A=L-S$ hence $A$ = $F$. As synchronous consensus is not possible with a larger $F$, it is also not possible with a larger $A$.
		
		
	\subsection{Leaderless Consensus}
	
		\textbf{Theorem 3}:	\textit{Leaderless consensus over a medium susceptible to asynchrony and faults.}
	
		Consensus algorithm, of a system of synchronous processes that communicate over a medium susceptible to asynchrony and faults, does not need to operate with a leader if it can ensure direct or indirect bounded delivery of messages among the required for consensus majority of correct processes.
		
		\textbf{Proof}
	
		Let us assume existence of a group $G$ of correct processes, equal or larger than the required for consensus quorum, with ensured bounded delivery, so that every message received by one process of $G$ is also received by every process of $G$ and included a group $M$ messages individually maintained by every process in $G$. Not included in $M$ are the messages: 1) directly received after expiry of 1 RTTB on processor's clock since the start of the current phase; 2) received over a 2-hop channel after expiry of 1.5 RTTB; and 3) received over a 3-hop channel after expiry of 2 RTTB intervals. 
		
		Thus, all processes in $G$ receive within a bounded time the same set of messages $M$, extract exactly the same data, and produce atomically consistent individual states. The consensus value computed by the processes of $G$ is produced from atomically consistent states, and used as input to instances of the same deterministic function. Hence the consensus values individually produced by the processes of $G$ are atomically consistent, as if proposed by a leader.
		
	\subsection{Ensured Both Safety and Bounded Liveness}

		\textbf{Theorem 4}: \textit{ Possibility for deterministic Byzantine consensus with conceptually ensured simultaneous safety and bounded liveness in partial synchrony.}
		
		A synchronous deterministic Byzantine consensus algorithm, capable to tolerate asynchrony and faults on its communication medium, operates with conceptually ensured simultaneous safety and bounded liveness in the area within its boundaries of tolerance.
		
		\textbf{Proof}
		
		Use of asymmetrically authenticated messages \cite{RivestShamirAdleman78} lets tolerate some  asynchronous / faulty links.
		It prevents a Byzantine process from affecting consensus safety [see Handling Byzantine Deceit in subsection 3.3 and Theorem 3 proof]. Theorem 1 proves bounded delivery within the boundaries of tolerance to links' asynchrony and/or faults and faulty processes [see Handling Byzantine Omission and Handling Byzantine Delay in subsection 3.3]. Theorem 3 proves that under bounded delivery consensus does not need leader. Having ensured leaderless consensus and bounded delivery in spite of links' asynchrony and/or faults and in spite of processes' crashes and/or omissions and/or delays, nothing capable to prevent bounded termination has left non-eliminated. Hence, within its boundaries of tolerance the algorithm operates with conceptually ensured simultaneous safety and bounded liveness.
		
		
	\section{Summary of the Results}
		
		This work demonstrated the following new features and qualities related to deterministic consensus algorithms in partial synchrony:
		
		1. A concept for proactive handling of asynchrony with virtual indirect multi-hop channels.
		
		2. Higher tolerance to links asynchrony than the leader-based algorithms with asymmetrically authenticated messages.
		
		3. Tolerance to any fault of any process, within the tolerated limits of faults and asynchrony, with no effect on bounded liveness.
		
		4. A fully decentralized deterministic Byzantine consensus algorithm with precisely defined tolerance to asynchronous links.
		
		5. Conceptually ensured bounded termination and conceptually ensured simultaneous safety and bounded liveness. 
		
		6. Simulator of consensus work, revealing events' permutations capable to prevent bounded liveness [see the Appendix].
		
		7. A methodology for regression analysis of simulation results and extrapolation [see the Appendix].
		
		8. Analytically defined tolerance to asynchrony in consensus systems per number of faulty processes [see the Appendix].
		
		9. Superlinear increase of the tolerated asynchronous links with every added pair of system processes [see the Appendix].
		
	\section{Conclusion}
		We presented a synchronous deterministic Byzantine consensus algorithm in partial synchrony. It operates with the weakest needs for communication synchrony, has no single point of failure, and ensures simultaneous safety and bounded liveness
		
		These features are outcome from a trade-off: the tolerance to faults and asynchrony is paid with limits on the efficient scalability and halved throughput. The gained non-existing at present features enable creation of new technologies with impact in areas ranging from financial services to cyber resilience.
		 
		The algorithm lets build technologies solving the obstructed efficiency of financial markets: in terms of time – with real-time settlement of trades; and in terms of consistency – with elimination of conditions allowing creation of phantom stocks. 
		
		A decentralized infrastructure built with the algorithm operates with enforced quantifiable cyber resilience.
		Multiple concurrent security breaches installing malicious code and multiple faulty and/or asynchronous links within the tolerated limits cannot disrupt it.

\begin{acks}
	This work is partly sponsored by the Australian Federal Government through the Research and Development Tax Incentive Scheme. 
	The author expresses his gratitude to Simeon Simoff 
	for his help with combinatorics, suggestions on formalization, and comments.
\end{acks}

\newpage

	\bibliographystyle{ACM-Reference-Format}
	\bibliography{SCDIS}
	
	\newpage

\appendix
	\section*{Appendix}
		
	\section{Boundaries of Tolerance}
		\textit{Note}: Throughout the next two sections, term 'correct link' denotes 'correct synchronous link' and term 'faulty link' denotes 'faulty or asynchronous link' for neatness of the presentation.
		
		Theorem 1 demonstrated that circumvention of faulty links with 3-hop indirect delivery is paid with an increase of time needed to establish agreement from 2 to 4 RTTB intervals. Paying that price requires knowledge of what exactly will be received in return. So, we need to find out the number of tolerated faulty links per $0, 1, 2, …, F$ faulty processes for every system of $2F+1$ processes.
		
		To formulate the goal, for a particular number of total processes we need to compute all permutations of faulty links and faulty processes per each tolerated number of faulty processes and test each computed permutation whether each process in a group of $F+1$ correct processes receives a message from every process in that group. The goal is to discover the highest number of faulty links per faulty processes where every permutation solves consensus.

	\subsection{The Simulator}
		
		Different permutations of the same number of correct links may need 2-hop or 3-hop indirect channels to solve consensus. With a system of 7 processes and a few faulty, the possible permutations with faulty processes and faulty links are millions or billions.
		
		Discovering the highest number of tolerated faulty links per faulty processes with solvable consensus requires a simulator to compute all permutations and test each one individually. The computation could happen on a single computer. A one-way link could be implemented as a shared memory, which can be modified by exactly one process and read by exactly one other process.

		With this objective, we built a simulator that distributes the algorithm's messages in a fully sequential manner, according to the permutation of correct processes and correct links. A completed distribution of messages is considered a completion of consensus round. Receiving a message by a recipient from a sender is considered receiving of both Phase One and Phase Two messages.
		
		Simulation was implemented according to the system model's
		common denominator for process faults, links asynchrony, and link faults. Omissions and/or delays of a Byzantine-faulty process are simulated in their worst-case scenario, where the max harm of a Byzantine-faulty process is exactly the same as with a stop-failed process. So, any form of faulty process is modeled as a stop-failed process. Asynchronous link is modeled as a faulty link, i.e. with value 0 in matrix $C$ \eqref{eq:1}. A stop-failed process is modeled with value 0 of anyone of its inbound and outbound links in matrix $C$ \eqref{eq:1}.

 	\begin{figure}[h]
 		\scalebox{0.10}
 		{\includegraphics{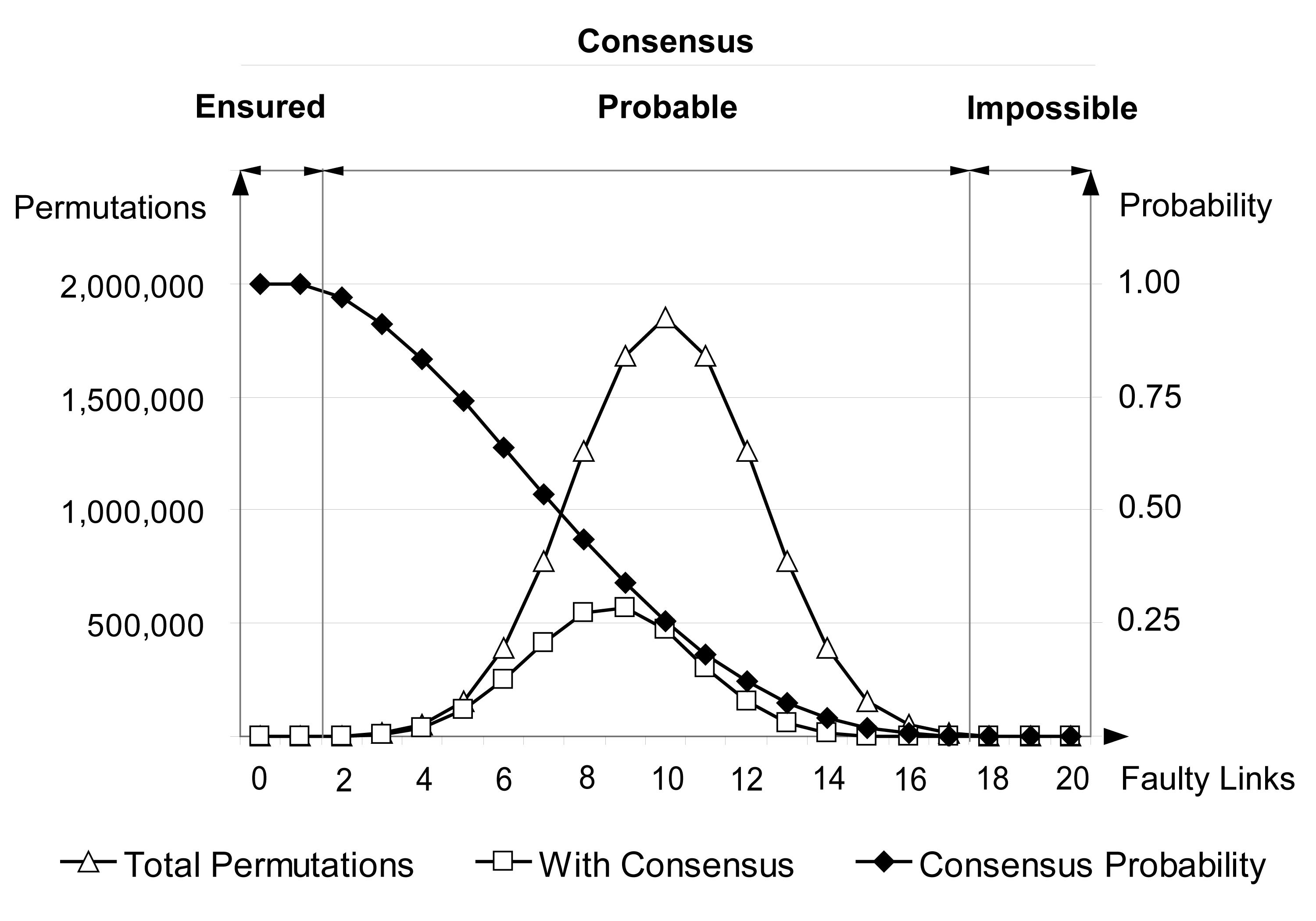}}
 		\caption{Consensus with 2 Faulty of 5 Processes}
 	\end{figure} 
 		
		On completed distribution of messages, the simulator checks for the existence of a group of $F+1$ correct processes, wherein every process of the group has received a message from every other process of that group. If yes, the consensus is considered solved. 

		The simulator operates in the following manner:

		\textbf{Input}: As input it takes: a) total number of processes; b) number of correct processes; and c) number of correct links.
		
		\textbf{Execution}: It creates objects representing processes and links, traces all permutations of correct processes and correct links, and checks each permutation for ability to solve consensus.
		
		\textbf{Output}: As output it returns: a) total number of permutations; and b) number of permutations allowing to solve consensus.
		
	\subsection{Computation with a 5-process System}
		
		According to the model, a 5-process system can tolerate up to 2 faulty (Byzantine or stop-failed) processes. We performed three series of computations, each comprising a series of 21 computations itself – the first with 0 faulty links and the last with 20. Every computation traces all permutations of faulty processes and faulty links and tests each permutation for ability to solve consensus. 
		
		\textbf{Series with Two Faulty Processes}
		
		Figure 2 depicts 3 diagrams presenting:
		
		- Total permutations, per number of faulty links;
		
		- Permutations that solve consensus, per number of faulty links; 
		
		- Probability to solve consensus, per number of faulty links. 
 	
 		In regard to probability (number of permutations that solve consensus divided by total permutations), Figure 2 shows 3 regions:
 		
 		- Region with \textbf{ensured} consensus, where probability is 1;
 		
 		- Region with \textbf{probable} consensus, where $0 < $ probability $< 1$; 
 		
 		- Region with \textbf{impossible} consensus, where probability is 0. 	
 		
 		As it is shown on Figure 2, a 5-process system operating with 2 faulty processes ensures consensus with up to 1 faulty link. In presence of 2 faulty links, the system solves consensus with 1,840 permutations out of 1,900 total, hence operates with 0.968 probability to solve it.
 	
 	\textbf{Series with One Faulty Process}
 	
 		Figure 3 also depicts 3 diagrams presenting: total permutations, permutations that solve consensus, and probability to solve consensus, per number of faulty links. In regard to probability, the major change is expansion of the region where consensus solving is ensured. 
 		
  	\begin{figure}[h]
 	\scalebox{0.10}
 	{\includegraphics{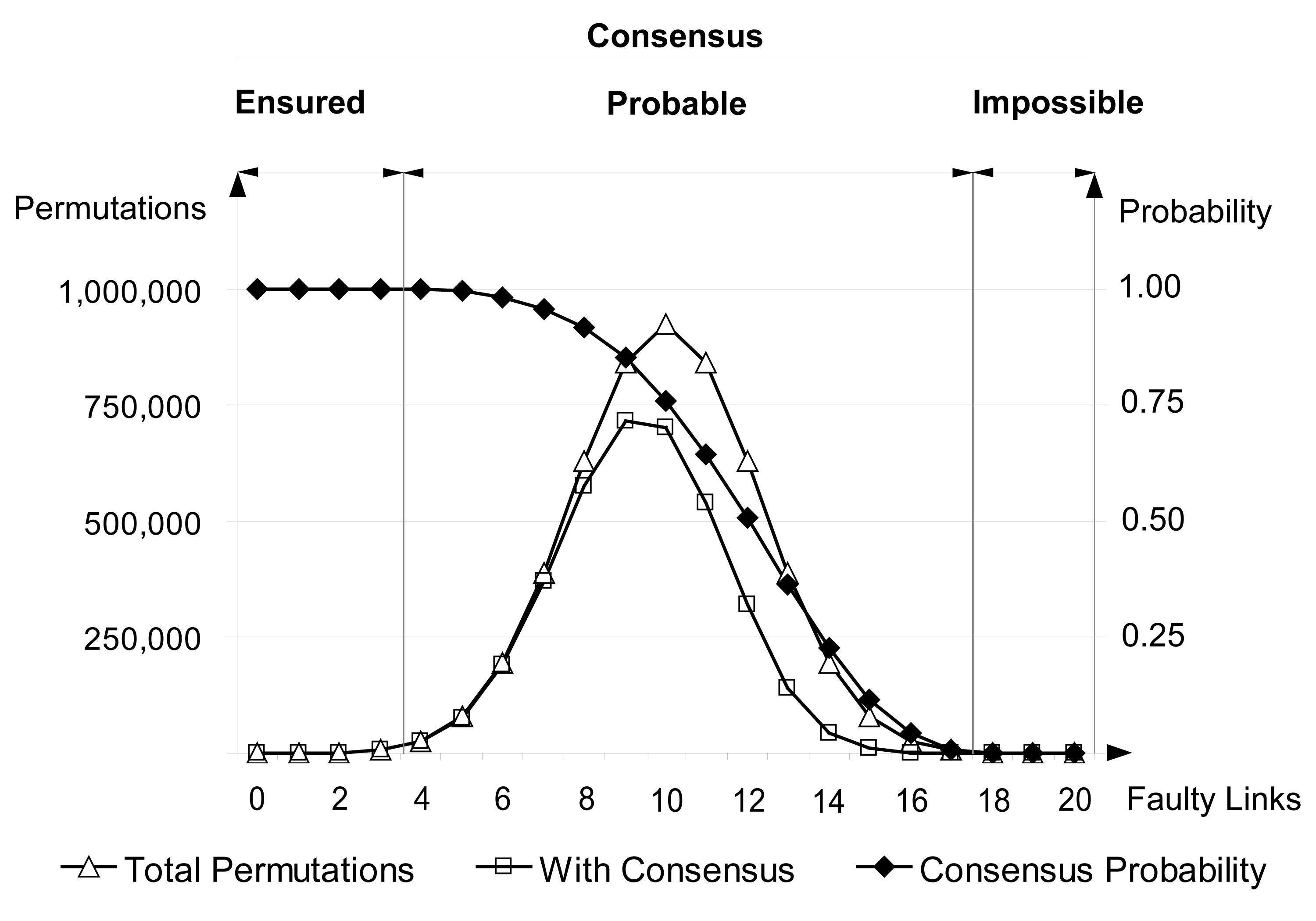}}
 	\caption{Consensus with 1 Faulty of 5 Processes}
	\end{figure}	
 		
 		This expansion has a simple explanation. Recall the modeling of a faulty process. All inbound and outbound links are dead with a stop-failed process or are considered dead as a worst-case scenario with a Byzantine process.
 		
 		As Figure 3 shows, a 5-process system operating with 1 faulty process ensures consensus with up to 3 faulty links. In presence of 4 faulty links, the system solves consensus with 24,195 permutations out of 24,225 total, hence operates with 0.9987 probability to solve it. In presence of 5 faulty links, the system solves consensus with 24,074 permutations out of 24,225 total, hence operates with 0.9938 probability to solve it.
 	
 	\textbf{Series with No Faulty Process}
 	
 		Figure 4 as well depicts the diagrams presenting: total permutations, permutations that solve consensus, and probability to solve consensus, per number of faulty links. In regard to probability, compared to the diagrams on the previous two figures, the major change again is the expansion of the region with ensured solving of consensus. Yet with no faulty processes the expansion is huge: from tolerance of 3 faulty links with 1 faulty process to to tolerance of 9 faulty links.
 	
 		As Figure 4 shows, a 5-process system operating with no faulty process ensures consensus with up to 9 faulty links. In presence of 10 faulty links, the system solves consensus with 184,696 permutations out of 184,756 total, hence operates with 0.9986 probability to solve it. In presence of 11 faulty links, the system solves consensus with 167,420 permutations out of 167,960 total, hence operates with 0.9967 probability to solve it. 
 	
 	\section{Challenge and Solution}
 	
 		Checking whether a consensus system of 9 processes can unconditionally tolerate simultaneous 3 faulty processes and 7 faulty links requires tracing and verifying for solvability of consensus 123 billion (more accurately 123,741,215,136) permutations, which took 58 days to complete the computation. The same system when operates with 2 faulty processes is expected to tolerate 11 faulty links. Checking for unconditional solvability of consensus requires tracing and verifying 108 trillion (more accurately 108,802,275,708,672) permutations, which requires 54,401 days, which is 149 years. Even after taking into consideration the symmetry factor of 9, the computation would have taken 16 years performed on a single computer. 
 
  	\begin{figure}[h]
 	\scalebox{0.10}
 	{\includegraphics{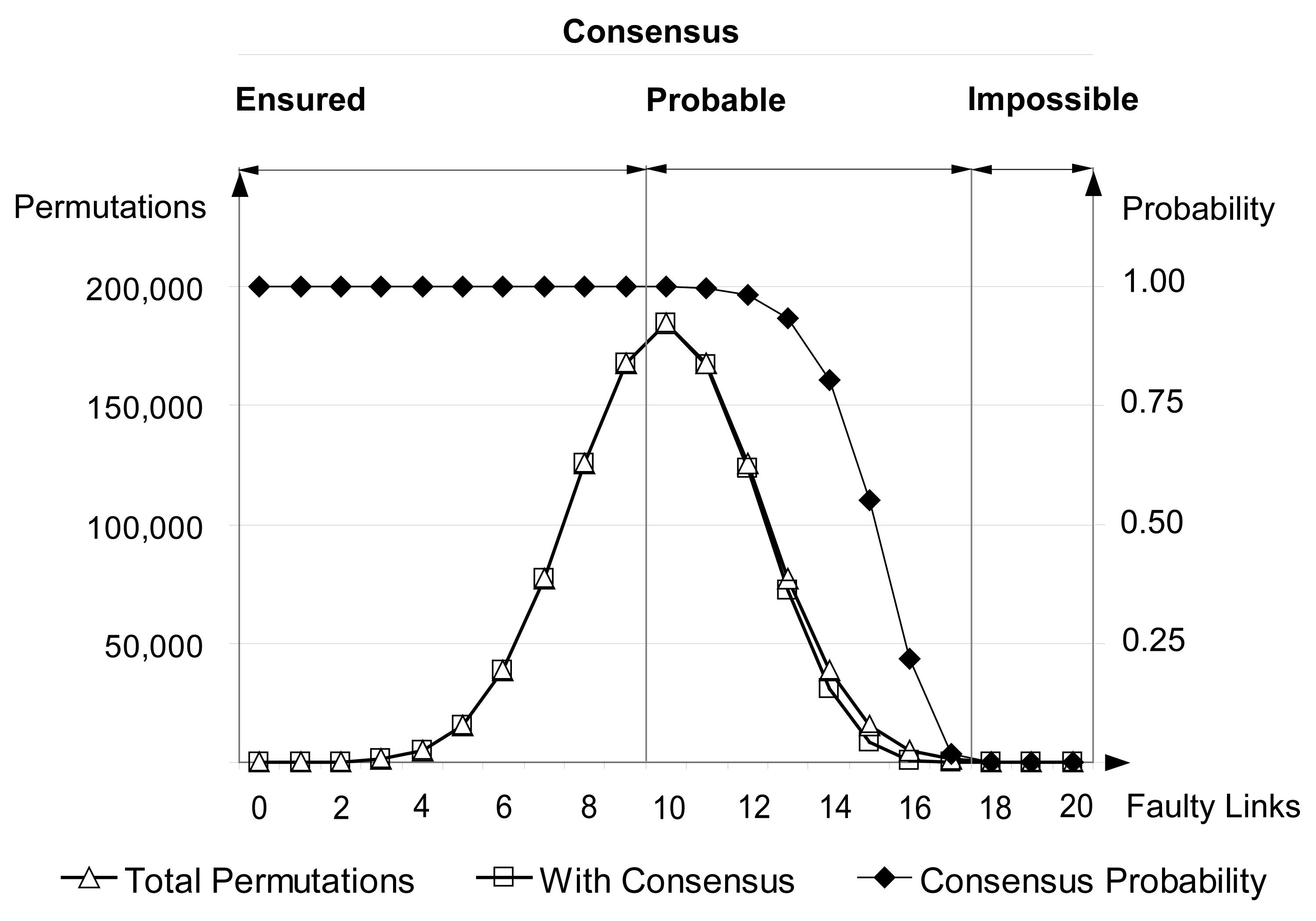}}
 	\caption{Consensus with 0 Faulty of 5 Processes}
 	\end{figure}
 		
 		Hence the ability of our simulation system to compute all permutations involving faulty processes and links, and to examine each one for ability to solve consensus is practically useless with a 9-process system or larger. 
 		Figure 5 presents the tolerance to faulty links of a 5-process, a 7-process, and a 9-process system, computed on multiple simulation systems within a time frame of 2 months.
 	
 	\subsection{Solution Idea}
 		
 		With a few exceptions, tolerance boundaries of a system of any size cannot be established with a mainstream computing technology. The alternative is to do it analytically in the following manner:
 		
 		- Extract as much data as possible with computation;
 		
 		- Use the computation data to fill a regression table;
 		
 		- Use the regression table to build tolerance equations;
 		
 		- Apply the equations to a systems of any size.
 		
 		Computations with a 6-process system with no faulty processes show that the system tolerates up to 4 faulty links. This means that under any permutation of 4 faulty links, the system unconditionally delivers all sent messages. The same system increases its tolerance to faulty links from 4 to 7 when it has to deliver messages of any 5 processes to the same 5 processes, and from 7 to 15 faulty links when it has to deliver from any 4 processes to the same 4 processes.
 		
 		\textbf{Creation of regression table}
 		
 		First step is to fill a row with the computed results showing how the tolerance to faulty links in delivery from $N$ to $N$ processes in an $N$-process system increases with an increment of $N$ with 1. Then to fill the next row with experimentally computed results showing how the tolerance to faulty links in delivery from $(N-1)$ to $(N-1)$ processes in an $N$-process system increases with an increment of $N$ with 1, then from $(N-2)$ to $(N-2)$ processes, etc. 
 		
 \begin{figure}[h]
 	\scalebox{0.10}
 	{\includegraphics{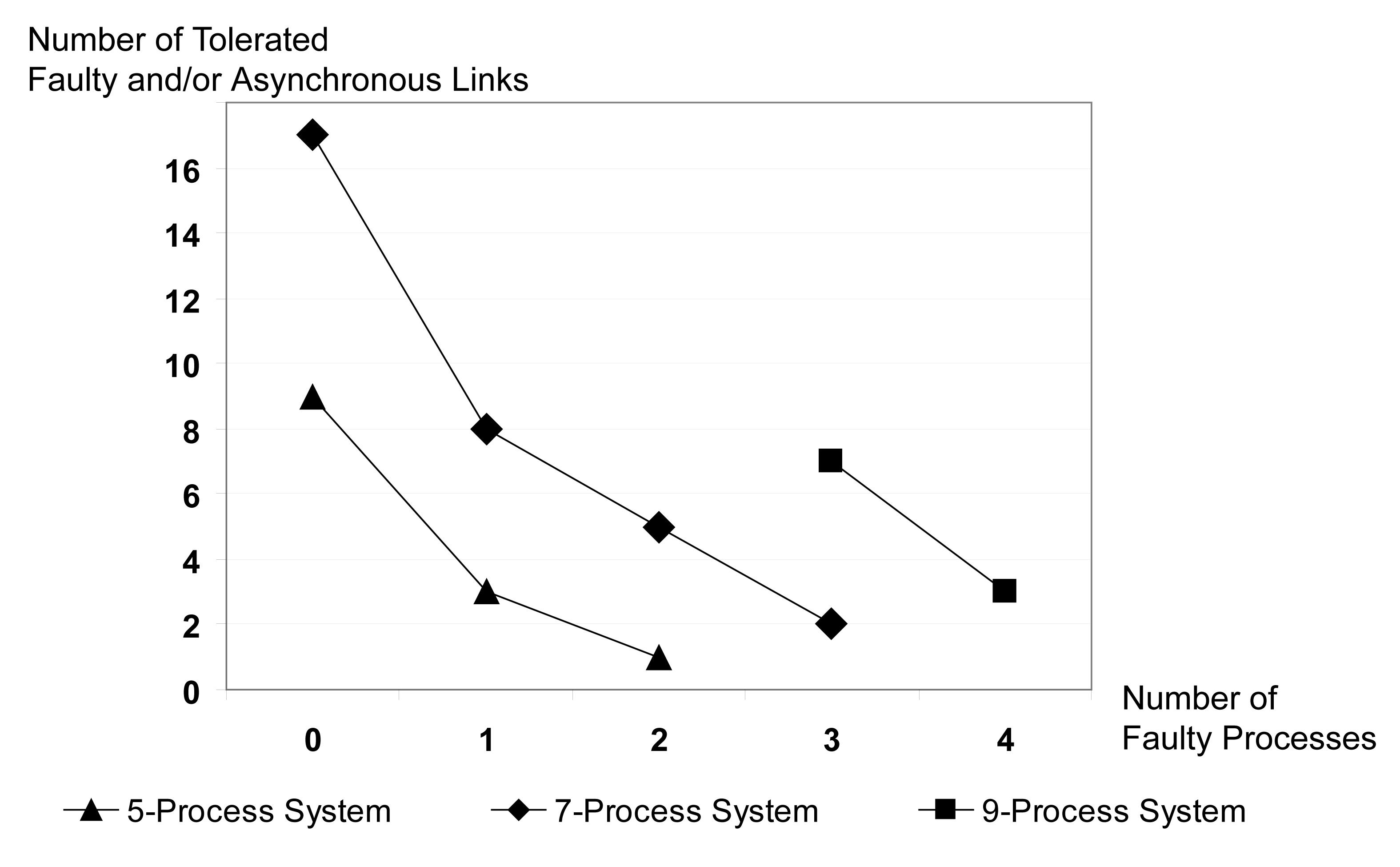}}
 	\caption{Tolerance to Async/Faults (Computation Results) }
 \end{figure}	
 		
 		\textbf{Illustration of the approach}
 		
 		Consider the tolerance to faulty links $f$ of the 7-process system on Figure 5 and the possible regression. The system tolerates up to $F=3$ faulty processes and number of $f$ faulty links, where with:
 		
 		- $F=3$ faulty processes, it is practically a 4-process system, which tolerates $f=3$ in delivering across the 4 correct processes.
 		
 		- $F=2$ faulty processes, it is practically a 5-process system, which tolerates $f=3$ in delivering across the 5 correct processes, and tolerates additional $+f=2$ faulty links in delivering across any 4 correct processes, so the resulting tolerance is $f=5$ faulty links.
 		
 		- $F=1$ faulty processes, it is practically a 6-process system, whose tolerance is equal to sum of tolerances in delivering across the 6 correct processes and the additional tolerance in delivering across any 5 correct processes and across any 4 correct processes.
 		
 	\subsection{Regression Table From Computations}
 	
 		Table 1 presents the results of computations on multiple simulator systems within a time limit set to 2 months per computation series. Individual computation series were performed with varied total number of processes  $N = 3, 4, 5, ..., 9$. Each one established tolerance to faulty links $f$ in delivery of $N$ messages across $N$ processes, $(N-1)$ messages across $(N-1)$ processes, etc.
 		
 		Table 1 data allows observation of the following 4 patterns:
 		
 		\textbf{Pattern 1}. In delivery of $N$ messages across $N$ processes, an increment of $N$ with 1 increases the faulty link tolerance $f$ with 1.
 		
 		\textbf{Pattern 2}. In delivery of $(N-1)$ messages across $(N-1)$ processes, starting from $N=4$, an increment of $N$ with 1 increments $\Delta$ with 1. The meaning of $\Delta$ is the additional tolerance ($+f$), caused by simplifying the objective by 1, i.e. from delivery of $N$ messages across $N$ processes to delivery of $(N-1)$ across $(N-1)$ processes. 
 		
 		\textbf{Pattern 3}. In delivery of $(N-2)$ messages across $(N-2)$ processes, starting from $N=6$ there is a repetition of Pattern 2. Hence, Pattern 3 can be formulated as: Starting from $N=4$, an increment of $N$ with 2 causes a new start of Pattern 2 in regard to a simplified by 1 objective. i.e. from delivery of $(N-1)$ messages across $(N-1)$ processes to delivery of $(N-2)$ messages across $(N-2)$ processes. 		
 		
 		\textbf{Pattern 4}. With $N-F=F+1$, where $F$ is the highest tolerated number of faulty processes, starting from $N=3$, an increment of $N$ with 2 causes an increment of $\Delta$ with 2. This pattern is clearly observed in the following sequence: 1)) N=3, delivery from $(N-1)$ to $(N-1)$, $\Delta=2$; 2) N=5, delivery from $(N-2)$ to $(N-2)$, $\Delta=4$; and 3) N=7, delivery from $(N-3)$ to $(N-3)$, $\Delta=6$.
 		
 \begin{table}[ht]
 	\caption{Computation Results}
 	\begin{tabular}{c c c c c c c c c}
 		\hline
 		Delivery Type &N 		&3&4&5& 6& 7& 8& 9 	\\
 		\hline
 		N to N 	  &$f$ 		&1&2&3& 4& 5& 6& 7 	\\
 		(N-1) to (N-1)&$\Delta$ &2&1&2& 3& 4& 5& 	\\
 		(N-1) to (N-1)&$f$	 	&3&3&5& 7& 9&11& 	\\
 		(N-2) to (N-2)&$\Delta$ & &5&4& 1& 2&  &	\\	
 		(N-2) to (N-2)&$f$		& &8&9& 8&11&  &	\\
 		(N-3) to (N-3)&$\Delta $& & & & 7& 6&  &	\\
 		(N-3) to (N-3)&$f$		& & & &15&17&  &	\\
 		\hline	\end{tabular}		
 \end{table}

 \begin{table}[ht]
 \caption{Computation and Extrapolation Results}
 \begin{tabular}{c c c c c c c c c c c}
 \hline
 Delivery Type &N 		&3&4&5& 6& 7& 8& 9&10&11 	\\
 \hline
  N to N 	   &$f$ 	&1&2&3& 4& 5& 6& 7& 8& 9 	\\
 (N-1) to (N-1)&$\Delta$&2&1&2& 3& 4& 5& 6& 7& 8 	\\
 (N-1) to (N-1)&$f$	 	&3&3&5& 7& 9&11&13&15&17 	\\
 (N-2) to (N-2)&$\Delta$& &5&4& 1& 2& 3& 4& 5& 6	\\	
 (N-2) to (N-2)&$f$		& &8&9& 8&11&14&17&20&23	\\
 (N-3) to (N-3)&$\Delta$& & & & 7& 6& 1& 2& 3& 4	\\
 (N-3) to (N-3)&$f$		& & & &15&17&15&19&23&27	\\
 (N-4) to (N-4)&$\Delta$& & & &  &  & 9& 8& 1& 2	\\	
 (N-4) to (N-4)&$f$		& & & &  &  &24&27&24&29	\\
 (N-5) to (N-5)&$\Delta$& & & &  &  &  &  &11&10	\\
 (N-5) to (N-5)&$f$		& & & &  &  &  &  &35&39	\\
 \hline
 \end{tabular}		
 \end{table} 
 
  	\subsection{Regression Table Analytically Expanded}	
 
 		Table 2 presents together the results from computation and from analytical extrapolation, based on the observed cause and effect patterns between the incremented number of system processes $N$ and the various $\Delta$s. Recall that
 		value of $f$ in a particular row and column is computed by adding the value of $\Delta$ in the previous row of the same column to the of preceding $f$ in that column.
 
 	\subsection{Examples: Use of Extrapolation}
 	
 		Tolerance of a 9-process system to faulty links per actual number of faulty processes $F$ can be computed (with $F=4$, $F=3$, and $F=2$) using data from Table 1 and (with $F=1$ and $F=0$) using data from Table 2 in the following manner:
 	
 		- With $F=4$, a 9-process system is equivalent to a 5-process system, which tolerates $f=3$ in delivering 5 messages to 5 processes (Table 1).
 	
 		- With $F=3$, the system is equivalent to a 6-process system, which tolerates $f=4$ in delivering 6 messages to 6 processes (Table 1), and tolerates additional $+f=3$ in delivering 5 message to 5 processes (Table 1), so the resulting tolerance is $f=7$.
 	
	 	- With $F=2$, the system is equivalent to a 7-process system, which tolerates $f=5$ in delivering 7 messages to 7 process (Table 1), tolerates additional $+f=4$ in delivering 6 messages to 6 processes (Table 1), and tolerates $+f=2$ in delivering 5 messages to 5 processes (Table 1), so the resulting tolerance is $f=11$.
 	
	 	- With $F=1$, the system is equivalent to a 8-process system, which tolerates $f=6$ in delivering 8 messages to 8 processes (Table 1), tolerates additional $+f=5$ in delivering 7 messages to 7 processes, tolerates additional $+f=3$ in delivering 6 messages to 6 processes (Table 2, Pattern 3), and tolerates additional $+f=1$ in delivering 5 messages to 5 processes (Table 2, Pattern 3), so the resulting tolerance is $f=15$.
 	
 	 \begin{figure}[h]
 		\scalebox{0.122}
 		{\includegraphics{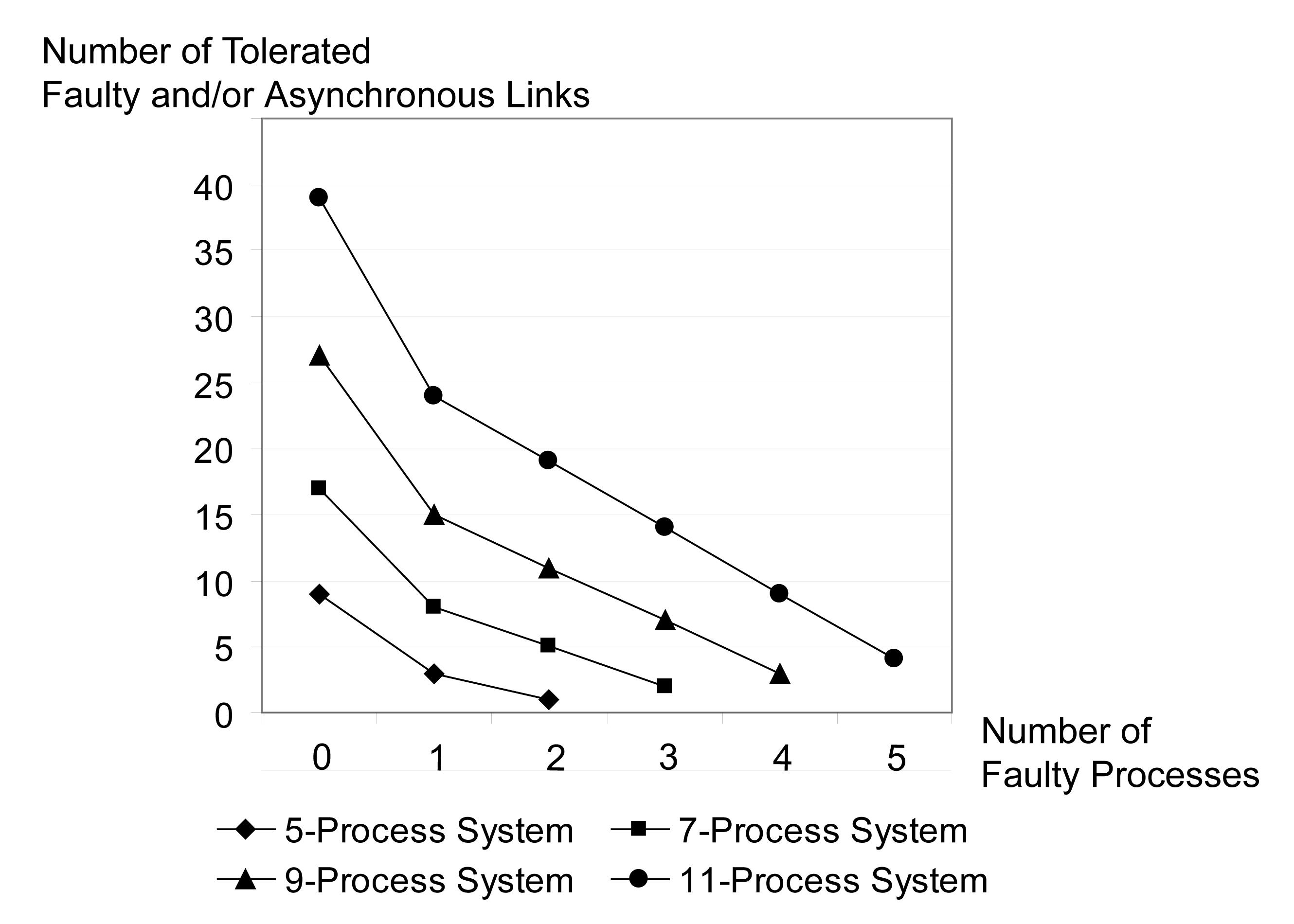}}
 		\caption{Tolerance to Async/Faults (Computed + Analytical) }
 	\end{figure}
 	
	 	- With $F=0$, the system is a 9-process system, which tolerates $f=7$ in delivering 9 messages to 9 processes (Table 1), tolerates additional $+f=6$ in delivering 8 messages to 8 processes (Table 2, Pattern 3), additional $+f=4$ in delivering 7 messages to 7 processes, additional $+f=2$ in delivering 6 messages to 6 processes(Table 2, Pattern 3), and additional $+f=8$ in delivering 5 messages to 5 processes (Table 2, Pattern 4), so the resulting tolerance is $f=27$.

	 	Tolerance of an 8-process system to faulty links per actual number of faulty processes $F$ can be computed (with $F=3$, $F=2$, and $F=1$) using data from Table 1 and (with $F=0$) using data from Table 2 in the following manner:
	 	
	 	- With $F=3$, an 8-process system is equivalent to a 5-process system, which tolerates $f=3$ in delivering 5 messages to 5 processes (Table 1).
	 	
	 	- With $F=2$, the system is equivalent to a 6-process system, which tolerates f=4 in delivering 6 messages to 6 processes (Table 1), and tolerates additional $+f=3$ in delivering 5 message to 5 processes (Table 1), so the resulting tolerance is $f=7$.
	 	
	 	- With $F=1$, the system is equivalent to a 7-process system, which tolerates $f=5$ in delivering 7 messages to 7 process (Table 1), tolerates additional $+f=4$ in delivering 6 messages to 6 processes (Table 1), and tolerates $+f=2$ in delivering 5 messages to 5 processes (Table 1), so the resulting tolerance is $f=11$.
	 	
	 	- With $F=0$, it is an 8-process system, which tolerates $f=6$ in delivering 8 messages to 8 processes (Table 1), tolerates additional $+f=5$ in delivering 7 messages to 7 processes (Table 1), tolerates additional $+f=3$ in delivering 6 messages to 6 processes (Table 2, Pattern 3), and tolerates additional $+f=1$ in delivering 5 messages to 5 processes (Table 2, Pattern 3), so the resulting tolerance is $f=15$.
	 	
	 	\textit{Note}: It might appear strange that an 8-process system with no faulty processes tolerates $f=15$, while a 7-process system with no faulty processes tolerates $f=17$. 
	 	This outcome illustrates why an additional process can make achieving an agreement harder, yet with no contribution to the tolerance to link faults or asynchrony. Consensus of an 8-process system requires delivery of 5 messages across 5 correct processes, while consensus of a 7-process system requires delivery of 4 messages across 4 correct processes.

 	\subsection{Tolerance Curves with Extrapolation}
	 	
	 	Figure 6 depicts a family of curves, where a curve represents the tolerance of a consensus system to faulty and/or asynchronous links per number of tolerated faulty processes. Tolerance of a 5-process system, a 7-process system, and a 9-process system with 4 and with 3 faulty processes is computed entirely with the simulator.

	 	Tolerance of a 9-process system with 2 faulty processes, with 1 faulty process, and with no faulty process is computed analytically with extrapolation. Tolerance of an 11-process system is computed entirely analytically with extrapolation.
		
	\subsection{Conditions for Bounded Termination}
 	
 		The following equations present tolerance of consensus system to the sum of faulty and asynchronous links $f$ as function of system connectivity $c$ (Graph theory meaning of term connectivity \cite{Deo1974}) and the number of faulty processes $F_{t}$ at a point of time $t$:		
 		
		- With odd total number of processes and no faulty processes:
		
	\begin{equation}
		\label{eq:8.1}
		\tag{8.1}
		\begin{Large}
			f < c + c/2 + (c/2)^2
		\end{Large}
 	\end{equation}
 	
 		- With odd total number of processes and $F_{t}$ faulty processes:
 		
	 \begin{equation}
	 	\label{eq:8.2}
	 	\tag{8.2}
	 	\begin{Large}
	 		f_{t} < c/2 + (c/2)^2 - F_{t}*c/2
	 	\end{Large}
	 \end{equation}	
 	 	
	 	- With even total number of processes and no faulty processes:
	 	
	 \begin{equation}
	 	\label{eq:8.3}
		\tag{8.3}
	 	\begin{Large}
	 		f < (c+1)/2 + ((c+1)/2)^2 - (c+1)/2
	 	\end{Large}
	 \end{equation}	
 
	 	- With even total number of processes and $F_{t}$ faulty processes:
	 	
	 \begin{equation}
	 	\label{eq:8.4}
	 	\tag{8.4}
	 	\begin{Large}
	 		f_{t} < (c+1)/2 + ((c+1)/2)^2 - (F_{t}+1)(c+1)/2
	 	\end{Large}
	\end{equation}
	
	These equations are the conditions for bounded termination.
	
\subsection{Final Comments}
	All results presented in sections B and C are obtained with a previous version of the algorithm, where 3-hop delivery between processes P and Q requires:
	1) P to have a correct synchronous outbound link to a process P' and a correct synchronous inbound link from the same process P';
	2) Q to have a correct synchronous outbound link to a process Q' and a correct synchronous inbound link from the same process Q'; and
	3) P' and Q' to be part of the correct majority and have correct synchronous links between themselves.
	
	The current version relaxes the 3-hop requirements. As Figure 1 shows,  delivery between processes P and Q requires:
	1) P to have a correct synchronous link to a process P' and a correct synchronous link from a process P'', where P' and P'' can be but not necessarily are the same process;
	2) Q to have a correct synchronous link to a process Q' and a correct synchronous link from a process Q'', where Q' and Q'' can be but not necessarily are the same process; and
	3) P', P'', Q', and Q'' to be part of the correct majority and have correct synchronous links between themselves.
	
	Hence the current version's tolerance to asynchronous and/or faulty links per number of faulty processes is \textbf{higher} than the presented here. Yet the actual borderline tolerance data is sensitive and kept confidential to prevent its use for fine tuning of coordinated cyber attacks on liveness of consensus networks that operate with the presented algorithm.

\end{document}